\documentclass[aps,pra,twocolumn,a4paper,amsmath,amssymb,showpacs,%
floatfix,]{revtex4}

\usepackage{graphicx}
\usepackage{epstopdf}

\newcommand{\Op}[1]{\boldsymbol{\mathsf{\hat{#1}}}}

\def\half{ \frac{1}{2}}
\def\openone{\leavevmode\hbox{\small1\kern-3.3pt\normalsize1}}

\begin{document}

\title{Vibrational stabilization of ultracold KRb molecules. A
  comparative study}
\date{\today}

\author{Mamadou Ndong}
\author{Christiane P. Koch}
\email{christiane.koch@physik.fu-berlin.de}
\affiliation{Institut f\"ur Theoretische Physik,
  Freie Universit\"at Berlin,
  Arnimallee 14, 14195 Berlin, Germany}
\date{\today}
\begin{abstract}
  The transfer of weakly bound KRb molecules from levels just below
  the dissociation threshold into the vibrational ground state with 
  shaped laser pulses 
  is studied. Optimal control theory is employed to calculate the
  pulses. The complexity of modelling the molecular structure is
  successively increased in order to study the effects of the
  long-range behavior of the excited state potential, resonant
  spin-orbit coupling and singlet-triplet mixing. 
\end{abstract}

\maketitle


\section{Introduction}
\label{sec:intro}

Research on cold and ultracold molecules has been one of the most active
areas of atomic and molecular physics over the last decade and continues
to draw much attention~\cite{ColdMolBook}.
Current activities are by and large still focussed on producing (ultra)cold
molecules, either by direct cooling or by assembling molecules from cooled
atoms using external fields. While direct methods such as
Stark deceleration~\cite{MeijerAnnuRevPhysChem06} 
have not yet reached the
regime of ultracold temperatures ($T\le 100\,\mu$K), photo- and
magneto-association create molecules in their electronic ground state
that are ultracold but in highly excited vibrational
levels~\cite{FrancoiseReview,JonesRMP06,KoehlerRMP06}.
However, prospective applications  ranging from high-precision
measurements to quantum information carriers~\cite{ColdMolBook}
require stable ultracold molecules. This has triggered the quest for
molecules in their absolute ground state. A major breakthrough toward
this long-standing goal was achieved when several groups produced
molecules in the lowest rovibrational level of an electronic ground
state potential via Stimulated Raman Adiabatic
passage~\cite{LangPRL08,NiSci08},
photoassociation~\cite{SagePRL05,DeiglmayrPRL08} and vibrational laser
cooling~\cite{PilletSci08}. Finally, the ability to control not only
the rovibrational but also the hyperfine degree of freedom has recently
paved the way toward ultracold molecules in their
absolute ground state~\cite{OspelkausPRL10}.

An earlier, alternative proposal to reach molecules in their vibronic
ground state 
was based on employing laser pulse shaping capabilities: An optimally
shaped laser pulse can coherently transfer, via many Raman transitions,
vibrationally highly excited molecules into $v=0$~\cite{MyPRA04}. The
attractiveness of this proposal rests on the fact that optimization is
carried out iteratively, both in experiment and in theory, and
does not require detailed knowledge of the molecular
structure to identify optimal pulses. One might nevertheless ask
whether and how \textit{qualitative} changes in the molecular
structure affect the optimal solution. This question defines the scope
of the present work. 
We solve the same optimization problem -- population transfer from a
weakly bound vibrational level just below the dissociation threshold
of the electronic ground state potential to $v=0$ -- for two 
different molecules, KRb and Na$_2$, successively taking effects into
account that strongly alter the molecular structure of KRb. Our choice 
to focus on this molecule is motivated by the
long-standing~\cite{RoatiPRL02,WangPRL04,ManciniPRL04,OspelkausPRL06,%
  NiSci08,KlemptPRA08,ThalheimerPRL08}
and continuing~\cite{OspelkausPRL10,AikawaNJP09,KimNJP09}
experimental efforts on KRb. 

\begin{figure*}[bt]
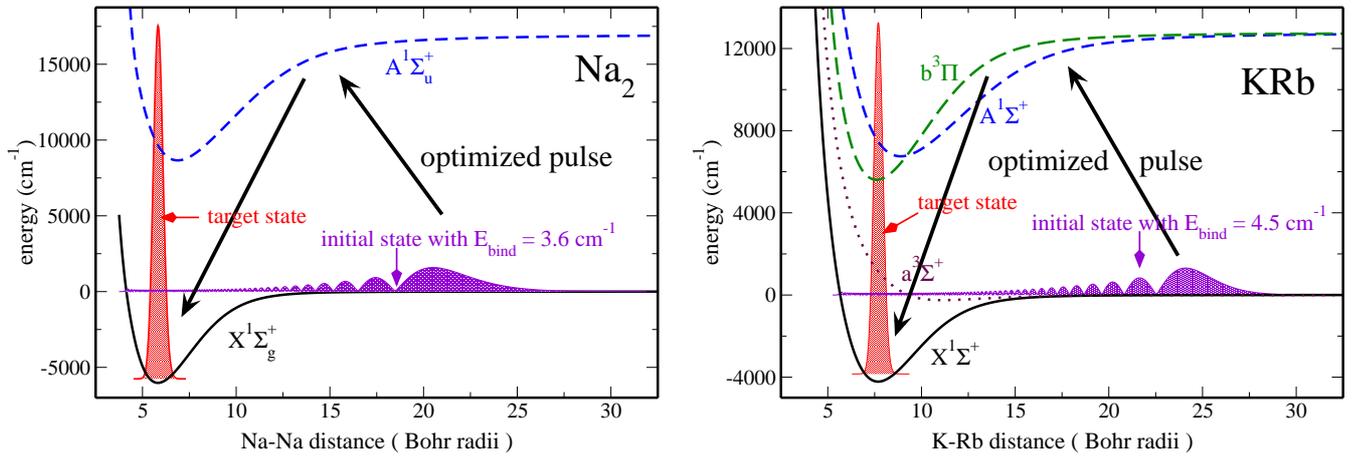

  \centering
  \includegraphics[width=0.48\linewidth]{figure1_1}\hspace*{3ex}%
  \includegraphics[width=0.48\linewidth]{figure1_2}  
  \caption{(color online) Potential energy curves and initial and
    target wave functions for vibrational stabilization of Na$_2$
    (left) and KRb (right) molecules. 
    The complexity of the model for the
    molecular structure of KRb is successively increased from two to
    four channels. 
  }
  \label{fig:scheme}
\end{figure*}

The paper is organized as follows: 
The general theoretical approach is described in
Section~\ref{sec:model}.
Sections~\ref{sec:longrange}-\ref{sec:mix} present and analyze the
short shaped laser pulses that achieve the vibrational transfer.
The complexity of the model for the molecular structure of KRb is
successively increased, cf. Fig.~\ref{fig:scheme}.
Based on a two-state model, Section~\ref{sec:longrange} compares the
optimization of vibrational transfer for KRb and
Na$_2$ molecules. They differ in the long-range behavior of their
excited state  potentials, $1/R^3$ vs $1/R^6$.
Section~\ref{sec:SO} is dedicated to the effect of
spin-orbit interaction in the electronically excited state which may
lead to resonant coupling between a singlet and a triplet
state~\cite{AmiotPRL99,HyewonMyPRA07,GhosalMyNJP09}. This is 
investigated with a three-state model in Section~\ref{sec:SO}.
Adding one more channel, 
Section~\ref{sec:mix} studies the transfer of a coherent superposition
of singlet and triplet molecules in their electronic ground state 
into the  ground vibrational level of the singlet potential.
This is possible for polar molecules due to the broken gerade-ungerade
symmetry~\cite{SagePRL05}. Section~\ref{sec:concl} concludes.  

\section{Theoretical approach}
\label{sec:model}

\subsection{Model}
\label{subsec:model}
The linear Schr\"odinger equation describing 
the internuclear dynamics of two atoms is considered. For 
molecules formed in a quantum degenerate gas, the many-body dynamics
are then neglected. This approach is justified by the time scales of
standard optical and/or magnetic traps~\cite{MyPRL09}.
While the internuclear dynamics and pulse shaping occur on the time scale of 
picoseconds, the many-body dynamics for conventional traps is
characterized by microseconds. The many-body system will have to adjust to the 
new internal state, but  this is going to happen on a much slower time scale
only after the pulse is over~\cite{PascalPRA03}.

The time-dependent Schr\"odinger equation, 
\begin{equation}
  \label{eq:Schroedinger}
  i\frac{\partial}{\partial t}|\psi(t)\rangle =
  \big(\Op{H}_0 + \Op{\mu}\varepsilon(t)\big) |\psi(t)\rangle
  \,,
\end{equation}
is solved with a Chebychev propagator \cite{RonnieReview94}.
The Hamiltonian comprises two or more electronic states as specified 
below, cf. Fig.~\ref{fig:scheme}. The interaction with the laser field
is treated in the dipole 
approximation. The rotating wave approximation is not invoked in
order to allow for strong fields and multi-photon transitions. 
The Hamiltonian and the wavefunctions,
$\psi_i(R)=\langle R|\psi_i\rangle$, where $i$ denotes the channel, are
represented on a Fourier grid with 
adaptive grid step \cite{SlavaJCP99,WillnerJCP04,ShimshonCPL06}. In
order to obtain vibrational eigenfunctions and Franck-Condon factors,
the Hamiltonian, $\Op{H}_0$, is diagonalized.

\subsection{Optimal control theory}
\label{subsec:OCT}

Denoting the formal solution of the Schr\"odinger equation at time $t$
by
\begin{equation}
  \label{eq:psi_t}
  |\psi(t)\rangle = \Op{U}(t,0)|\psi(0)\rangle\,,
\end{equation}
the objective functional for a transition from initial state
$|\psi_{ini}\rangle$ to target state $|\psi_{target}\rangle$ at the
final time $T$ can be written
\begin{equation}
  \label{eq:F}
  F = \big| \langle \psi_{ini} |\Op{U}^{\dagger}(T,0;\varepsilon) |\psi_{target}\rangle\big|^2\,.
\end{equation}
It corresponds to the overlap of the initial state that has been
propagated to time $T$ under the action of the field $\varepsilon(t)$
with the target state.
A field is optimal if it completely transfers the initial
state, $|\psi_{ini}\rangle$, to the target state, $|\psi_{target}\rangle$,
i.e. if $F$ reaches a value close to one.

The objective $F$ is a functional of the field $\varepsilon(t)$. It
explicitly depends only on the final time $T$. In order to
use dynamical information from intermediate times, a new functional is
defined,
\begin{equation}
  \label{eq:J}
  J = -F + \int_0^T g(\varepsilon) dt \,,
\end{equation}
where $g(\varepsilon)$ denotes an additional constraint over the
field. Often $g(\varepsilon)$ is chosen to minimize the pulse
fluence. This implies a replacement rule in the control equation for
the field. However, a choice of
$g(\varepsilon)$ that leads to vanishing 
changes in the field as the optimum is reached may be advantageous
from the point of view of convergence~\cite{JosePRA03}. It is employed here,
\begin{equation}
  \label{eq:g}
  g(\varepsilon)=\frac{\lambda}{S(t)}
  \bigg[\varepsilon(t)-\tilde\varepsilon(t)\bigg]^2\,,
\end{equation}
and $\tilde\varepsilon(t)$ is taken to be the field of the previous
iteration. This choice of $g(\varepsilon)$ implies an update rule rather
than a replacement rule in
the control equation for the field. Physically, it corresponds to
minimizing the change in pulse energy at each iteration.
The shape function $S(t)$, $S(t)=\sin^2(\pi t/T)$, enforces a smooth
switch-on and off of the field. The parameter $\lambda$ controls the
optimization strategy: A small value results in a small weight of the
constraint, Eq.~(\ref{eq:g}), allowing for large modifications in the
field, while a large value of $\lambda$ represents a conservative
search strategy with only small modifications in the field at each
iteration.

We employ the Krotov algorithm
\cite{KonnovAC99,SklarzPRA02,JosePRA03,KrotovDok08,KrotovAC09}  to obtain the
control equations. The derivation of the algorithm for the target,
Eq.~(\ref{eq:F}), and the constraint, Eq.~(\ref{eq:g}), is described
in detail in Ref.~\cite{MyPRA04}.
It yields the following prescription to improve the field,
\begin{widetext}
  \begin{equation}
    \label{eq:eps}
    \varepsilon_j(t) = \varepsilon_{j-1}(t) +
    \frac{S(t)}{\lambda}\mathfrak{Im}\left\{
        \langle \psi_{ini}|\Op{U}^{\dagger}(T,0;\varepsilon_{j-1})|\psi_{target}\rangle
        \langle \psi_{target}|\Op{U}^{\dagger}(t,T;\varepsilon_{j-1})\;\Op{\mu}\;
        \Op{U}(t,0;\varepsilon_{j})|\psi_{ini}\rangle
        \right\}
  \end{equation}
\end{widetext}
at the $j$th iteration step. The first overlap in the parenthesis can be
shown to be time-independent. The second overlap contains a backward
propagation of the target state from time $T$ to time $t$ under the old
field, $\varepsilon_{j-1}(t)$, and a forward propagation of the initial
state under the new field, $\varepsilon_{j}(t)$. 
Eq.~(\ref{eq:eps}) is implicit in $\varepsilon_{j}(t)$. This is
remedied by employing two different grids in the time discretization,
see Ref.~\cite{MyPRA04} for details. 

\section{The role of the long-range behavior of the excited state 
  potential: Comparing KRb to Na$_2$}
\label{sec:longrange}

The investigation is started with a simple two-state model. The
Hamiltonian reads
\begin{equation}
  \label{eq:H2surf}
  \Op{H}_{2s} (t)=
  \begin{pmatrix}
    \Op{T}  + V_{X^1\Sigma^+_{(g)}}(\Op{R})  & \Op{\mu}\, \varepsilon(t) \\ 
    \Op{\mu}\, \varepsilon^*(t) &  \Op{T}+V_{A^1\Sigma^+_{(u)}}(\Op{R})
  \end{pmatrix}\,,
\end{equation}
where $(g/u)$ applies only to Na$_2$.
$\Op{T}$ denotes the kinetic energy operator and $V_i(\Op{R})$ the
potential energy curve of channel $i$. For the Na$_2$
molecule, we employ the same
potential energy curves as in Ref.~\cite{MyPRA04}%
\footnote{Please note that an incorrect value for the 3S-3P excitation
energy of sodium was considered in Ref.~\cite{MyPRA04}. All spectra need to
be shifted by $5991\,$cm${-1}$. The conclusions of Ref.~\cite{MyPRA04} with
respect to spectral bandwidth, required optimization time and pulse
energy remain unchanged.} 
which were obtained from molecular
spectroscopy~\cite{SamuelisPRA00,TiemannZfPD96}. 
The ground state potential for KRb is also known
spectroscopically~\cite{PashovPRA07}. The excited state potential
energy curves of KRb are unfortunately not yet known with
spectroscopic precision. We have therefore employed the results of 
\textit{ab initio} calculations~\cite{RousseauJMolSpec00}
at short internuclear distances together with an asymptotic expansion
of the form $V_\textsf{asy}(R) = -\frac{C_6}{R^6}-\frac{C_8}{R^8}$
with the long-range coefficients taken from
Ref.~\cite{MarinescuPRA99}.
The laser field,
$\varepsilon(t)$, couples to the molecule via the transition
dipole moment, $\Op{\mu}$. The latter is expected to depend on
internuclear distance at least for short $R$. However, it is
approximated by $\Op{\mu}=1\,$a.u. since no data on the $R$-dependence
could be found in the literature.

The initial state, $\psi_{ini}(R)=\langle R|\psi_{ini}\rangle=\psi_g^v(R)$
is taken to be a vibrational wavefunction of the electronic ground
state corresponding to a weakly bound level just below threshold.
Counting the levels downward from the last bound one, $v=v_{last}-3$ for
Na$_2$ and $v=v_{last}-6$ for KRb. These levels were chosen to yield a
comparable binding energy for the two species. They are very similar to
levels that would be populated when molecules are formed from atoms
using a Feshbach resonance or photoassociation but with a 
somewhat larger binding energy.
This approximation eases calculations since, as discussed below, the
absolute value of the optimization time is determined by the binding
energy of the initial state. It does not affect the comparison between
Na$_2$ and KRb which is based on relative times.

While the model, Eq.~(\ref{eq:H2surf}),
does not capture the full complexity of the molecular
structure, it allows for a straightforward comparison with earlier
work on the Na$_2$ molecule \cite{MyPRA04}. In particular, it is
useful to highlight the influence of the long-range behavior of the excited
state potential which is $1/R^6$ for KRb but $1/R^3$ for Na$_2$. 
From a spectroscopic point of view, one expects the $1/R^6$-behavior
of KRb to be more favorable for vibronic transitions between highly
excited vibrational levels. This is
attributed to better Franck-Condon overlaps between alike potentials. 
However, from a dynamical point of view, a maximum difference
potential between electronic ground and excited state, $\Delta
V(R)=\half(V_e(R)-V_g(R))$, is more desirable. A larger difference
potential is obtained for an excited state with
$1/R^3$ long-range behavior, i.e. for Na$_2$.
The reasoning behind this argument is that 
the wave packet that launched by the pulse on the excited state
experiences a much larger gradient and can better
accelerate its motion toward shorter distances~\cite{KallushPRA07}.

In order to decide which of the two arguments is relevant for
vibrational stabilization of ultracold molecules, we determine the
minimum pulse energy and minimum optimization time required to achieve
a transfer to the target state of 99\% or better. This is based on the
fact that many solutions to the control problem exist. Which solution
will be found by the algorithm depends crucially on the boundary
conditions -- optimization time and pulse power or pulse
energy. However, if the resources in terms of time and energy are
insufficient, no solution will be found. The lower limits to $T$ and
$\mathcal{E}_P$ can thus be used to characterize the control problem
and the solution strategy.

\begin{figure}[tb]
  \centering
  \includegraphics[width=0.99\linewidth]{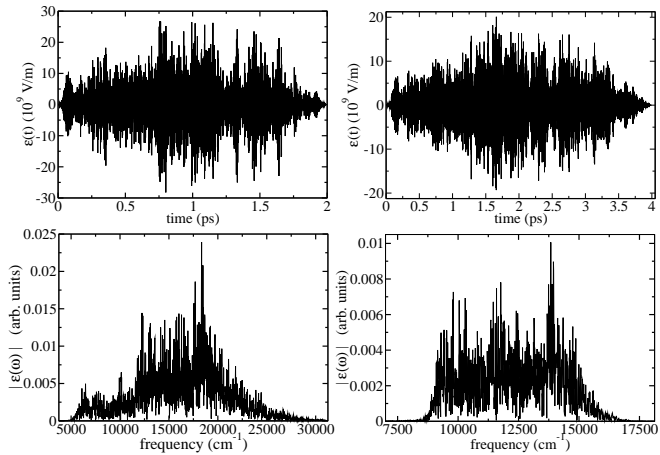}
  \caption{Optimal pulses (upper panel) and their spectra (lower
    panel) for Na$_2$ (left) and KRb (right).}
  \label{fig:optfield1}
\end{figure}

Figure~\ref{fig:optfield1} compares the optimal pulses and their
spectra for Na$_2$ (left) and KRb (right).
These pulses are the results of a three-step optimization. Initially, an
optimization time of $T=16\,$ps corresponding to about twice the vibrational
period of the initial state of Na$_2$ was chosen~\cite{MyPRA04}. The
guess pulses for this optimization were constructed as a series of
100$\,$fs pulses with two central frequencies reflecting the peaks of
the Franck-Condon factors of the initial and the target states.
This choice provides a large enough spectral bandwidth for given $T$.
The pulses were optimized until a transfer of 99\% or better was
achieved. For the second optimization step, the pulses were compressed
in time following the recipe detailed in Ref.~\cite{MyPRA04},
i.e. points $\varepsilon(\omega_i)$ 
were removed such that $\Delta\omega$ is increased and $T$
decreased. The resulting pulses were then employed as guess pulses for
the second step of optimization.
For Na$_2$, compression of $T$ by a factor of 8 was possible while for KRb a
factor of 4 turned out to be the limit. For larger compression factors,
the optimization did not result in any appreciable population transfer to the
target state. The difference in the minimum optimization time $T$ for
Na$_2$ and KRb is explained in terms of the different reduced masses
of the molecules. Although the binding energy of the initial states is
comparable,  cf. Fig.~\ref{fig:scheme}, the motion of the heavier KRb
is slower. This is reflected in the vibrational period of the initial
states, 8$\,$ps for Na$_2$ compared to 14$\,$ps for KRb.
Therefore,  the compression factor taken with respect to the
corresponding vibrational period, is 4 for Na$_2$ compared to 3.5 for
KRb, i.e. the maximum factor for compression in time is very similar.
Finally, the minimum energy required for optimal transfer was
determined in step three where the guess pulses were taken to be 
the optimal pulse of the previous step divided in amplitude by some
factor. If the factor was too large, optimization did not result in
any appreciable population transfer, otherwise 99\% transfer or better
were achieved. This way, a sharp limit for the mininum required pulse
energy was obtained. It amounts to 78$\,\mu$J for Na$_2$ and 61$\,\mu$J for
KRb where the focal radius of the laser beam is assumed to be
100$\,\mu$m.

The lower bounds on the optimization time and the pulse energy
represent the main difference between the optimal pulses for Na$_2$
and KRb. As can be seen in Fig.~\ref{fig:optfield1}, the overall
temporal and spectral structures of the pulses are fairly similar.
The spectral bandwidth of the optimal pulse is significantly 
larger for Na$_2$ than for KRb. It is, however, difficult to attach
a physical meaning to the spectral bandwidth. We do not filter
out undesired spectral components~\cite{LapertPRA09,SchroederNJP09},
i.e. those components that do not correspond to given vibronic
transitions in order to avoid significantly increased convergence
times and numerical effort. A constraint formulated directly in
frequency domain to contain the bandwidth within a certain spectral
region cannot be enforced simultaneously with
Eq.~\eqref{eq:g}~\cite{MyPRA04}. Therefore,  with progressing
iteration, the spectral bandwidth of the optimized pulses grows. Since
in principle this growth can be suppressed by
filtering~\cite{LapertPRA09,SchroederNJP09}, it is rather an artifact
of the algorithm than a physically significant finding.

\begin{figure}[tb]
  \centering
  \includegraphics[width=0.98\linewidth]{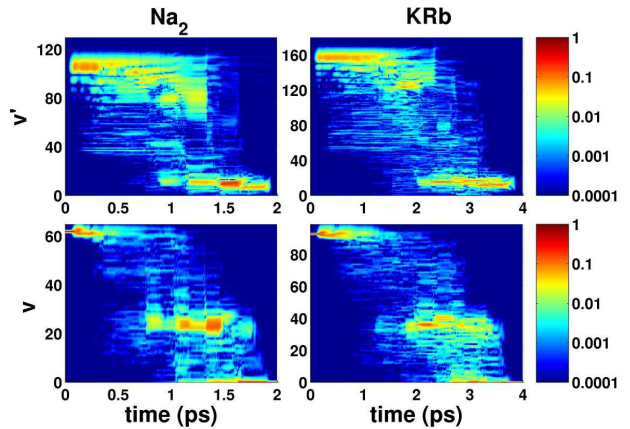}
  \caption{(color online)
    Projection of the time-dependent wave packets onto the vibrational
    eigenstates of the electronic ground (bottom) and
    excited (top) states for Na$_2$ (left) and KRb (right)} 
  \label{fig:pop1}
\end{figure}
\begin{figure}[tb]
  \centering
  \includegraphics[width=0.98\linewidth]{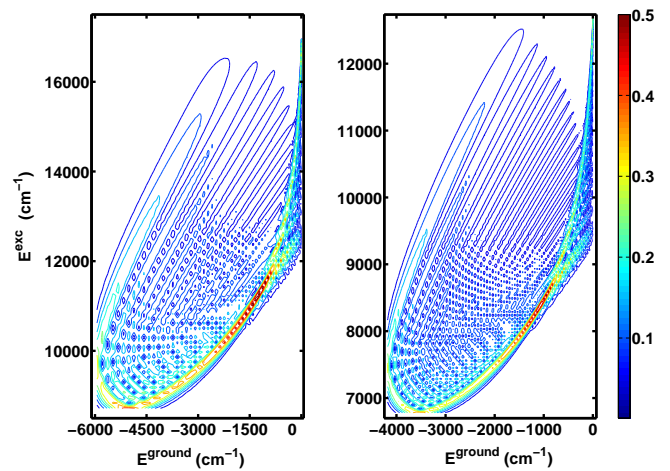}
  \caption{(color online)
    Franck-Condon factors of the ground state levels with all
    excited state levels for Na$_2$ (left) and KRb (right).
    The dissociation limit of the electronic ground state defines
    the zero of energy and the 
    dissociation limit of the excited state is $16965\,$cm$^{-1}$ for Na$_2$ 
    and  $12737\,$cm$^{-1}$ for KRb.}
  \label{fig:fcf1}
\end{figure}
The dynamics induced by the optimal pulses are analysed in
Fig.~\ref{fig:pop1} by projecting the ground and excited state
components of the time-dependent wave packets onto the vibrational
eigenstates. The dynamics show similar features for Na$_2$ and KRb. The
optimal pulse depletes the initial state and pumps most of the
population to the electronically excited state, distributing it over a
wide range of highly excited vibrational levels.
The wave packet then climbs down the potentials. In the middle of the
optimization time interval, it reaches ground state levels
$v\approx 30$ ($v\approx 40$)
for Na$_2$ (KRb) ground state which is about half way down to the
bottom of the potential well. In the last quarter of the optimization
time interval, population is accumulated in a superposition of a few
excited state levels around $v' \approx 10$ for Na$_2$ ($v' \approx 18$
for KRb). In a final step, this superposition is transfered to the
target level, $v=0$.

The population dynamics is rationalized by an analysis of the
Franck-Condon map, cf. Fig.~\ref{fig:fcf1}. The dynamics start
in the upper right corner of the map and follow the main ridge until
$v\approx 30$ ($v\approx 40$) for Na$_2$ (KRb) is reached. At this
point the Franck-Condon map takes approximately the shape of a
parabola where the right branch connects to $v\approx 30$ ($v\approx
40$)  Na$_2$ (KRb) while the left branch connects to the target level,
$v=0$. This explains the dynamics in the second half of the
optimization time interval where the population is pumped into the
excited state levels, $v' \approx 10$ for Na$_2$ ($v' \approx 18$
for KRb), that are reached by the two branches of the parabola,
i.e. that are the ideal gateway to $v=0$.

To conclude the comparison of Na$_2$ and KRb,
the minimum optimization time is dictated by the mass
of the molecule and population transfer in terms of required pulse
energy is more favorable for a $1/R^6$ than a $1/R^3$ excited state
potential. The overall dynamics of the population transfer is rather
similar for the two molecules and is easily rationalized by the
structure of the Franck-Condon map.

\section{Spin-orbit coupling in the electronically excited state}
\label{sec:SO}

The complexity of our model is increased to take spin-orbit
interaction in the electronically excited state of KRb into
account. Similarly 
to the Rb$_2$ and RbCs molecules, spin-orbit interaction may lead to
resonant coupling and strong non-adiabatic effects
\cite{AmiotPRL99,HyewonMyPRA07,GhosalMyNJP09}. 
This is captured by a three-state Hamiltonian,
\begin{widetext}
\begin{equation}
  \label{eq:H3surf}
  \Op{H}_{3s} (t) =
    \begin{pmatrix}
    \Op{T}  + V_{X^1\Sigma^+}(\Op{R})  & \Op{\mu}\, \varepsilon(t) & 0 \\ 
    \Op{\mu}\, \varepsilon^*(t) &  \Op{T}+V_{A^1\Sigma^+}(\Op{R}) &
    W_{SO}^{\Sigma\Pi}(\Op{R})\\
    0 & W_{SO}^{\Sigma\Pi}(\Op{R}) & \Op{T}+V_{b^3\Pi}(\Op{R}) - W_{SO}^{\Pi\Pi}(\Op{R})
  \end{pmatrix}\,,
\end{equation}
\end{widetext}
where the diagonal and off-diagonal spin-orbit interaction terms,
$W_{SO}^{\Pi\Pi}(\Op{R})$ and $W_{SO}^{\Sigma\Pi}(\Op{R})$, are
introduced. 
In principle, the $b^3\Pi$ excited state has a dipole coupling with
the lowest triplet state. One can thus transfer molecules from the
lowest triplet state via Raman transitions into the singlet ground
state. This will be investigated below in Section~\ref{sec:mix}, and
the present section is devoted to studying the effect of
non-adiabaticities in the electronically excited state. Here, the
initial state is purely singlet, i.e. the same weakly vibrational level
of the $X^1\Sigma^+$ state, $v=93=v_{last}-6$, as in the previous section is
considered. 

Unlike in the case of Rb$_2$ where the spin-orbit interaction terms were
determined spectroscopically~\cite{BergemanJPB06}, no such accurate
data is available for KRb. We have therefore resorted to the
parametrization of $W_{SO}^{\Pi\Pi}(\Op{R})$ and
$W_{SO}^{\Sigma\Pi}(\Op{R})$ in terms of Morse functions,
\begin{eqnarray}
  \label{eq:W_SO}
  W_{SO}^j(\Op{R}) &=& P_1^j + \left(P_2^j-P_1^j\right)\left(
    1 - e^{P_3^j\left(P_4^j-\Op{R}\right)} \right)^2\,,\\
  \nonumber &&\quad\quad j=\Pi\Pi,\Sigma\Pi\;,   
\end{eqnarray}
that was
introduced by Bergeman et al. for RbCs~\cite{BergemanPRA03}. 
These two functions show a dip at intermediate distances and level
off toward a constant value at long range.
As a first guess, we have employed the values for the parameters
$P_i^j$ from Ref.~\cite{BergemanPRA03} scaled 
to reproduce the correct asymptotic limit of the
fine-structure splitting of 237.595$\,$cm$^{-1}$. The corresponding
values for the parameters $P^j_i$ are listed in the first row of
Table~\ref{tab:SO}. 
\begin{table*}[tb]
  \centering
  \begin{tabular}{|c|c|c|c|c|c|c|c|c|c|c|}
    \hline
    & j & $P^j_1$ (cm$^{-1}$) & $P^j_2$  (cm$^{-1}$) & $P^j_3$ (\AA$^{-1}$) & $P^j_4$ (\AA) & j & $P^j_1$ (cm$^{-1}$) & $P^j_2$  (cm$^{-1}$) & $P^j_3$ (\AA$^{-1}$) & $P^j_4$ (\AA) \\ \hline
    RbCs \cite{BergemanPRA03}, scaled
    & $\Pi\Pi$ &135.49 & 184.70 &0.24 & 5.82 & $\Sigma\Pi$&130.77 &261.20 &0.23 & 5.85 \\ \hline
    case 1 & $\Pi\Pi$ &180.49 & 184.70 & 0.24 & 5.82 & $\Sigma\Pi$&130.77  & 261.20 & 0.5 & 5.85 \\ \hline
    case 2 & $\Pi\Pi$ &135.49 & 184.70 & 0.24 & 5.82 & $\Sigma\Pi$& 130.77 & 261.20 &  0.5 & 5.85 \\     
    \hline
  \end{tabular}
  \caption{Parameters of the spin-orbit coupling functions,
    cf. Eq.~\eqref{eq:W_SO}.} 
  \label{tab:SO}
\end{table*}
Since the parameters $P_i^j$ are not accurately
known, we have varied the $P_i^j$ in order to estimate the maximum
effect that the spin-orbit interaction can have on the vibrational
wave functions and Franck-Condon matrix elements.
This provides the starting point for
studying the  strongest possible effect of the spin-orbit interaction
on the optimization and the dynamics under the optimal pulse.
Two different choices of spin-orbit coupling are employed, referred to
below as cases 1 and 2. The
corresponding parameters are listed in the second and third rows of
Table~\ref{tab:SO}. The modification of the parameters is quite
substantial and larger than what can realistically be
expected. However, the point here is to demonstrate the most positive
and most negative effect that the spin-orbit coupling may have on the
vibrational stabilization and to explore its influence on the
optimization. 

In case 1, we have modified $P_1^{\Pi\Pi}$ and
$P_3^{\Sigma\Pi}$. The latter corresponds to  the width of the
dip in the off-diagonal spin-orbit coupling, while the former represents 
the constant offset of the diagonal spin-orbit coupling, which essentially
causes a relative shift of the vibrational ladders of the $A^1\Sigma$ and
$b^3\Pi$ states. This choice of parameters
leads to strong resonant coupling and strongly
perturbed vibrational wavefunctions where each diabatic component
shows peaks at the four classical turning points of both potentials.
As illustrated in the middle panel of Fig.\ref{fig:psi}, such a
situation is potentially favorable for vibrational stabilization:
\begin{figure}[tb]
  \centering
  \includegraphics[width=0.98\linewidth]{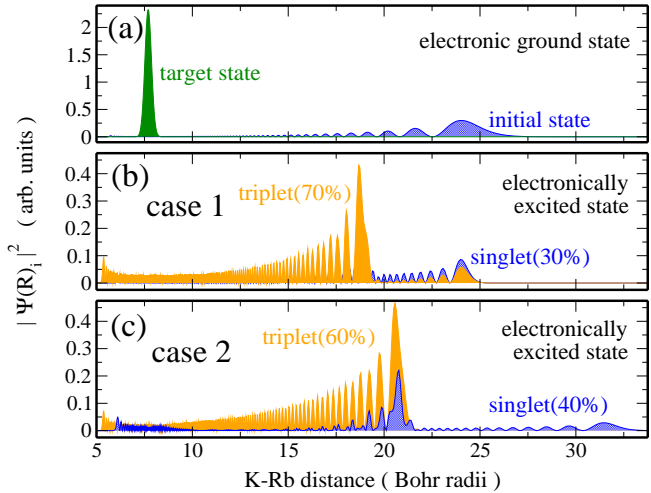}
  \caption{(color online)
    Initial and target vibrational wavefunctions of the electronic
    ground state (upper panel) and 
    effect of the spin-orbit coupling on the vibrational wavefunctions
    of the electronically excited states (medium and bottom panel).}
  \label{fig:psi}
\end{figure}
The outermost peak of the vibrational wavefunction with a binding
energy of $112\,$cm$^{-1}$ (corresponding to an absolute energy 
of $12625\,$cm$^{-1}$) 
leads to good Franck-Condon overlap with the
initial state. The singlet component of the wavefunction shows a
second peak at the outer turning of the 
$b^3\Pi$ state, $R\approx 19\,$a$_0$.
This second peak will lead to better overlap with more
deeply bound levels in the electronic ground state and could thus
cause a speed-up of the stabilization dynamics toward shorter
distances or less required pulse energy. Note that 
the Franck-Condon overlap reflects only the singlet component of the
vibrational wavefunctions. However, due to the time-dependence of the
stabilization process, the triplet component may play a role as
well. If there is a dynamical interplay of pulse and spin-orbit coupling, the
transfer efficiency can be much larger than predicted by static 
Franck-Condon overlaps~\cite{MyPRA06b}. Such a situation occurs for
strong pulses and
pulse durations comparable to or longer than the period
of the singlet-triplet oscillations caused by the spin-orbit
interaction. Both conditions will be met by 
the optimized pulses presented below.

In case 2, we have only modified the width of the dip in the off-diagonal
spin-orbit coupling, $P_3^{\Sigma\Pi}$. As in case 1, strong
perturbations in the vibrational wavefunctions are observed, cf. the
peaks at the two outer turning points in the singlet component in 
the lower panel of Fig.~\ref{fig:psi}. However, the 
spin-orbit coupling is now expected to have a detrimental effect on
vibrational stabilization where the wave packet shall be transferred
from large to short distances: Once the wavepacket comes close to the
outer turning point of the upper adiabatic potential,
$R\approx 21\,$a$_0$, the resonant spin-orbit coupling will move part
of the probability amplitude all the way out to the outer turning
point of the lower adiabatic potential, $R\approx 32\,$a$_0$. 
Therefore in case 2, the spin-orbit coupling will potentially
counteract the vibrational stabilization. 

\begin{figure}[tb]
  \centering
  \includegraphics[width=0.99\linewidth]{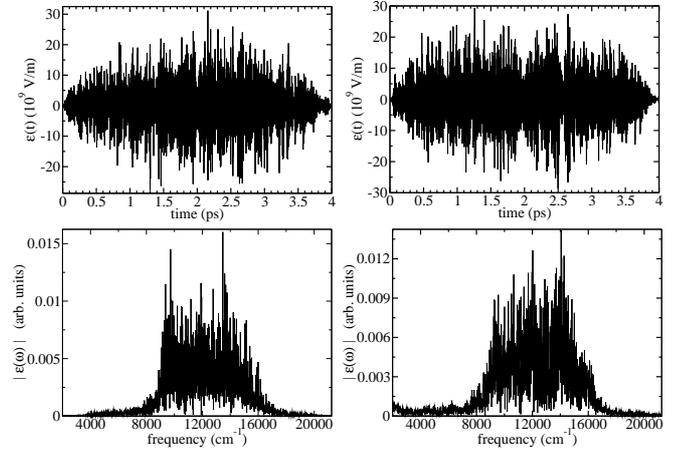}
  \caption{Optimal pulses (upper panel) and their spectra (lower
    panel) for the spin-orbit coupling cases 1 (left) and 2 (right).}
  \label{fig:optfield2}
\end{figure}

The same three-step optimization procedure as in
Section~\ref{sec:longrange} has been followed: (i) optimization for
$T=16\,$ps, (ii) compression in time to $T=4\,$ps and subsequent
re-optimization, (iii) determination of the minimal integrated pulse
energy with which a population transfer of better than 99\% can be
achieved. However, the guess pulses for step (i) were chosen such as
to take the modified Franck-Condon factors into account.
Figure~\ref{fig:optfield2} compares the optimal pulses and their
spectra for spin-orbit coupling cases 1 and 2.
Compared to Fig.~\ref{fig:optfield1} where the spin-orbit coupling in
the excited state was completely neglected, the optimal spectra of
Fig.~\ref{fig:optfield2} are broader, with additional spectral
amplitude at small and large frequency components. However, as
explained in Section~\ref{sec:longrange}, further calculations
employing spectral filtering are necessary to determine whether these
spectral features are artifacts of the optimization algorithm or
whether they represent a true physical requirement that the optimal
pulse has to fulfill. As seen in Fig.~\ref{fig:optfield2}, the minimum
optimization time to yield  a population transfer of better than 99\%
is not affected by the spin-orbit coupling in the excited state.
This is in accordance with
the rationalization in terms of the time scale of the vibrational
dynamics on the electronic ground state, i.e. in terms of the
timescales related to resolving the initial state and the target
state, cf. Section~\ref{sec:longrange}. 
The mininum required pulse energy amounts to 140$\,\mu$J for
spin-orbit coupling case 1 and 180$\,\mu$J for case 2. While the
optimization target, i.e. population transfer of better than 99\%,
can be achieved in both cases,
case 2 which had been identified as potentially bad for the
stabilization, 
requires more pulse energy. Both cases require substantially more
pulse energy than the estimate of $61\,\mu$J obtained with the
two-state model of Section~\ref{sec:longrange}. This is most likely
due to the much larger state space that is explored by the
optimization. 

\begin{figure}[tb]
  \centering
  \includegraphics[width=0.98\linewidth]{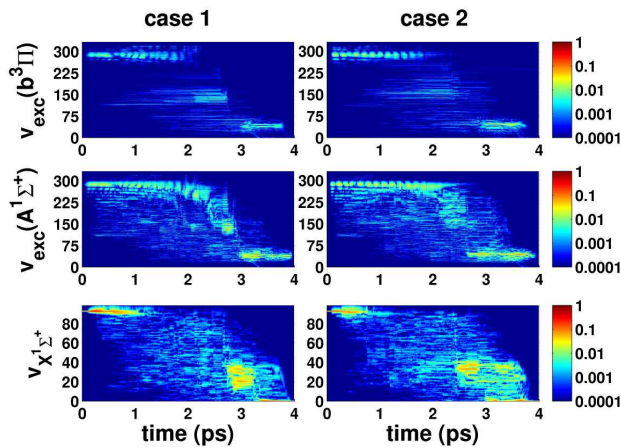}
  \caption{(color online)
    Projection of the time-dependent wave packets onto the vibrational
    eigenstates of the electronic ground (bottom) and the coupled 
    excited (top) states for spin-orbit coupling cases 1 (left) and 2
    (right).} 
  \label{fig:pop2}
\end{figure}
\begin{figure}[tb]
  \centering
  \includegraphics[width=0.98\linewidth]{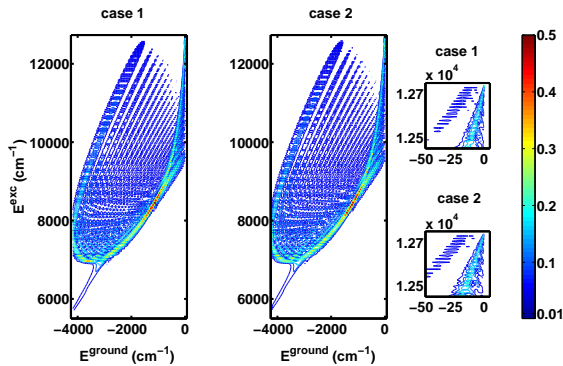}
  \caption{(color online)
    Franck-Condon factors of the ground state levels with the singlet
    component of the 
    excited state levels for Na$_2$ (left) and KRb (right).}
  \label{fig:fcf2}
\end{figure}
The dynamics under the optimal pulses are analyzed in
Fig.~\ref{fig:pop2} by projecting the time-dependent wave packet
onto the triplet and singlet components of the vibrational
eigenstates. Overall the dynamics are very similar for the two
spin-orbit coupling cases. A difference would be expected mainly at
the beginning of the pulse where the initial state is excited into
levels of about $12630\,$cm$^{-1}$ energy of the electronically
excited state. Inspection of the Franck-Condon map displayed in
Fig.~\ref{fig:fcf2} (inset) shows that the resonant coupling leads to 
additional peaks compared to the model without spin-orbit
interaction. These features are 
caused by the additional peak in the singlet component of the
vibrational eigenfunctions near the outer turning point of the triplet
potential, cf. Fig.~\ref{fig:psi}. However, the projections of the
wave packet within the first 1$\,$ps do not reveal any substantial
differences between coupling case 1 and 2. We therefore conclude that
the modifications of the Franck-Condon map due to the spin-orbit
interaction are not significant enough to influence the optimized
stabilization dynamics, no matter whether the type of  coupling is
potentially favorable or potentially detrimental. The complete
stabilization dynamics is rationalized in terms of the Franck-Condon
map analogously to Section~\ref{sec:longrange}, i.e. it is determined
by the main ridges of the Franck-Condon map. The only difference
between spin-orbit coupling cases 1 and 2 that can clearly be
identified  is the spread of population over the vibrational levels
which is larger for case 2.

To summarize the investigation of the influence of the spin-orbit
interaction in the electronically excited state, 
the minimum optimization time is not affected while the required pulse
energy is significantly increased compared to the model without
spin-orbit interaction. Details of the spin-orbit interaction have 
only a minor effect on the required pulse energy and stabilization
dynamics. 

\section{Optimizing transfer from a singlet-triplet superposition to
  the singlet ground state}
\label{sec:mix}

In heavy heteronuclear alkali dimer molecules it is possible to
transfer a vibrationally excited state 
that is in the lowest triplet state or in a superposition of the
lowest triplet and singlet electronic ground state, to the rovibronic
ground state~\cite{SagePRL05}. In order to study this as an optimization problem, 
a four-state model of the KRb molecule is considered,
\begin{widetext}
\begin{equation}
  \label{eq:H4surf}
  \Op{H}_{4s} (t) =
    \begin{pmatrix}
    \Op{T}  + V_{X^1\Sigma^+}(\Op{R})&0  & \Op{\mu}\, \varepsilon_\pi(t) & 0
    \\
    0 & \Op{T}  + V_{a^3\Sigma^+}(\Op{R})& 0 &  \Op{\mu}\, \varepsilon_\sigma(t)\\
    \Op{\mu}\, \varepsilon_\pi^*(t) &  0 & \Op{T}+V_{A^1\Sigma}(\Op{R}) &
    W_{SO}^{\Sigma\Pi}(\Op{R})\\
    0 & \Op{\mu}\, \varepsilon_\sigma^*(t) & W_{SO}^{\Sigma\Pi}(\Op{R}) &
    \Op{T}+V_{b^3\Pi}(\Op{R}) - W_{SO}^{\Pi\Pi}(\Op{R}) 
  \end{pmatrix}\,,
\end{equation}
\end{widetext}
that allows for transfer of molecules from the lowest
triplet state to the singlet ground state due to 
the spin-orbit interaction in the electronically excited state. 
The lowest triplet state potential is taken from
Ref.~\cite{PashovPRA07}, the other potential curves, dipole moments
and spin-orbit coupling functions are constructed as described in the
previous sections. In particular, the spin-orbit coupling cases 1
and 2 introduced in Section~\ref{sec:SO} are employed.
The initial state is taken to be a superposition of
the vibrational eigenfunctions of the $a^3\Sigma^+$
lowest triplet state and $X^1\Sigma^+$ singlet electronic ground
state with 4.5$\,$cm$^{-1}$ binding energy. The triplet (singlet)
component carries  70\% (30\%) of the population. The target state
remains unchanged compared to the previous sections, i.e. the $v=0$
level of the $X^1\Sigma^+$ singlet electronic ground state.

Different laser polarizations need to be taken into account -- linearly polarized
light, $\varepsilon_\pi(t)$, for the singlet transitions and
circularly polarized light, $\varepsilon_\sigma(t)$, for the triplet
transitions. This simply means that instead of Eq.~\eqref{eq:eps} two
equations for the two components of the field need to be considered
where the dot products are evaluated for the corresponding components
of the states.

As explained above, a three-step optimization procedure is carried out
in order to determine the minimum optimization time and minimum pulse
energies. Also for the four-state model,
population transfer with an efficiency of better
than 99\% is achieved by the optimal pulses. While the optimization
time remains unchanged at 4$\,$ps, the required pulse energy is
increased. It amounts to $270\,\mu$J for $\pi$-polarization
and $230\,\mu$J for $\sigma$-polarization in spin-orbit coupling case
1, and to $300\,\mu$J for $\pi$-polarization and $270\,\mu$J for
$\sigma$-polarization in case 2. As in Section~\ref{sec:SO}, case 1
which is potentially favorable for the stabilization requires slightly
less pulse energy than case 2. However, the further increase
of pulse energy compared to the three-state model of
Section~\ref{sec:SO} where pulse energies of 140$\,\mu$J and
180$\,\mu$J were obtained, is significant, in particular in view of
the fact that the molecule now couples to two polarization
components. The high pulse energies reflect the more difficult
optimization problem that is considered here where the 
wave function needs to be changed qualitatively from a
singlet-triplet superposition to a pure singlet state. 

\begin{figure}[tb]
  \centering
  \includegraphics[width=0.98\linewidth]{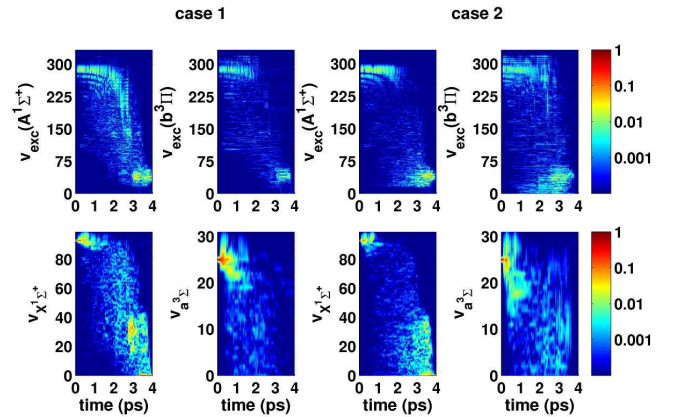}
  \caption{(color online)
    Projection of the time-dependent wave packets onto the vibrational
    eigenstates of the electronic ground (bottom) and the coupled 
    excited (top) states for spin-orbit coupling cases 1 (left) and 2
    (right).} 
  \label{fig:pop3}
\end{figure}
\begin{figure}[tb]
  \centering
  \includegraphics[width=0.98\linewidth]{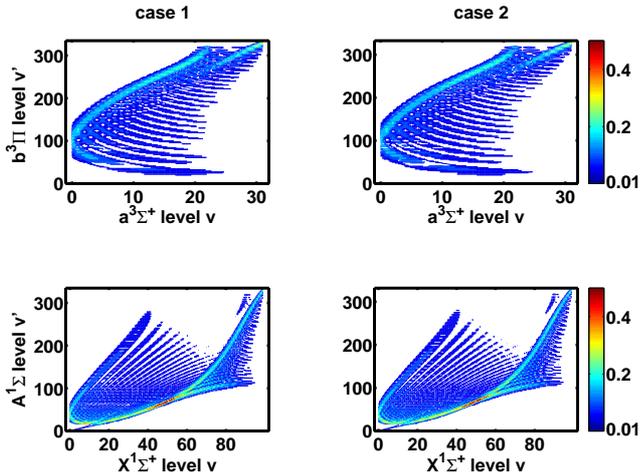}
  \caption{(color online)
    Franck-Condon factors of the singlet $X$ ground state levels with the singlet
    component of the  excited state levels and of the triplet $a$
    ground state levels with triplet component  of the  excited state
    levels for spin-orbit coupling cases 1 (left) and 2 (right).}
  \label{fig:fcf3}
\end{figure}
The dynamics under the optimal pulses are analyzed in terms of the
projections of the time-dependent wave packet onto the components of
the eigenfunctions on the four electronic states in Fig.~\ref{fig:pop3}. 
The dynamics of the singlet components is rather similar to those of
the previous sections, cf. Figs.~\ref{fig:pop1} and
\ref{fig:pop2}. Most of the triplet component of the initial state is
converted to singlet components within the first half of the pulse, in
particular in spin-orbit coupling case 1. In the potentially detrimental
case 2, more population resides in the triplet components than in case
1 and remains there throughout the pulse. 
The population dynamics should be compared to the Franck-Condon maps
shown in Fig.~\ref{fig:fcf3}. The singlet dynamics start out in the upper
right corner of the Franck-Condon maps. They follow the main ridges of
the map but fan out as the dynamics roll down the ridge and levels
with equally likele transitions to many levels are populated. The
triplet dynamics follows the right-most ridge in the upper right
corner of the Franck-Condon maps. It never jumps over to the left-most
ridge; and in particular in case 1, the range between $v'=275$ and
$v'=50$ of the $b^3\Pi$ state seems to be bridged via singlet
dynamics. This is an indication for
cooperative behavior between pulse and spin-orbit coupling in order to achieve
the triplet-singlet transfer required by the optimization task: The
dynamics travel down the right-most ridge of the Franck-Condon map due
to Rabi cycling, i.e. due to interaction with the circularly polarized
pulse. Then the spin-orbit coupling transfers most of the triplet
population to the singlet channel, and a further decrease of the
vibrational excitation happens in the singlet channels due to
interaction with the $\pi$-polarized light. Such a cooperative 
behavior can be expected since the optimization time is much larger than the
time scale corresponding to the spin-orbit interaction. In the
potentially detrimental case 2, no indication for cooperative behavior
between pulse and singlet-triplet oscillations is observed; and
population that was initially in the triplet channels remains there
for a much longer time.

To summarize this section, vibrational stabilization from a
singlet-triplet superposition 
to a pure singlet state can be achieved with better than 99\%
efficiency. However, even more pulse energy for both polarization
components is required compared to stabilization of a pure singlet
state. Depending on the details of the spin-orbit coupling function,
cooperative behavior between pulse and spin-orbit coupling may or may
not be observed.

\section{Conclusions}
\label{sec:concl}
We have studied the vibrational transfer of KRb molecules from a level
just below the dissociation limit to the vibrational ground
state. Optimal control theory was employed to obtain the shaped laser
pulses that drive this population transfer with an efficiency of 99\%
or better. As the main result, our
calculations have yielded an estimate of the minimum time that is
required for the vibrational transfer, i.e. the quantum speed limit
for this process~\cite{CanevaPRL10}, and an estimate on the required pulse
energy. 

Our findings have confirmed that optimal control approaches
work as 'black-box' algorithms
that provide solutions independent of the details of a given quantum
object. Nevertheless, our results cannot straightforwardly be transferred
to a laser pulse shaping experiment on cold molecules.
Insufficient knowledge of the molecular structure and restriction of
our model to a few electronic states prevent direct experimental
application of the calculated pulses. This is a common phenomenon
encountered in the optimal control of complex quantum systems such as
molecules~\cite{HornungJCP01} as opposed to atoms~\cite{BayerPRL09} or spin
systems~\cite{SkinnerJMR03}. In principle, our model could be refined. For
example, Ref.~\cite{MyPRA04} discusses in
detail how potential loss channels such as multiphoton ionization
could be incorporated; and more spectroscopy could be performed to
obtained a better knowledge of the potential curves and non-adiabatic
couplings. However, such refinement would miss the point of this
study. Here, we come to a two-fold 
conclusion: On one hand, our study encourages optimal control experiments
because solutions will be found no matter what are the specific
details of the molecule. On the other hand, our study has clarified,
as discussed below, 
the influence of the molecular structure relevant for the vibrational
transfer by successively increasing the complexity of the
model. While of less importance in optimal control
experiments based on feedback loops, these findings are important for
vibrational transfer and vibrational cooling using cw lasers,
incoherent broadband light or adiabatic passage.

First, we have addressed the role of the long-range behavior of
the excited state potential on the vibrational stabiliziation by
comparing the KRb and  Na$_2$ molecules. From a time-dependent
perspective one might expect the $1/R^3$ potential of the homonuclear
sodium dimer to be more favorable for the vibrational transfer since
the larger slope of the potential speeds up the motion toward shorter
internuclear distances. However, from a time-independent perspective, one might
argue that the $1/R^6$ excited state potential of 
heteronuclear molecules yields better Franck-Condon overlap with the
electronic ground state that also shows a $1/R^6$ dependence at large
internuclear distances. Comparing a two-state model, i.e. a model
comprising of the singlet electronic ground state and a single excited
state, for Na$_2$ and  KRb, we found that significantly less pulse
energy is required for KRb. We therefore conclude that a $1/R^6$
excited state potential is more favorable for vibrational transfer
than a $1/R^3$ potential, i.e. the spectroscopic perspective prevails
over the dynamical one.

The comparison of the KRb and  Na$_2$ molecules has also allowed us to
identify what determines the minimum  optimization time, i.e.  the
quantum speed limit for vibrational stabilization. The longest
timescale in the problem that needs to be resolved by the optimal
pulse is the vibrational motion of the initial state. Taking
comparable binding energies of the initial state for KRb and  Na$_2$,
the difference in the vibrational periods is due to the mass of the
molecules, and the minimum  optimization time is smaller for the
lighter molecule.

Second, we have increased the complexity of the model for the molecular
structure by taking spin-orbit coupling in the electronically excited
state into account. The resulting model consists of three electronic
states where the two electronically excited states exhibit a
non-adiabatic coupling.
In heavy alkali dimer molecules, the spin-orbit interaction does not
only modify the potentials at large internuclear 
separation, it may also cause a mixing of vibrational ladders
affecting the complete vibrational spectrum. This effect has been
termed 'resonant coupling'~\cite{AmiotPRL99,HyewonMyPRA07,GhosalMyNJP09}.
Since the $R$-dependence of the spin-orbit coupling function for KRb
is not precisely known, we have adapted a parametrization developed by
Bergeman et al. for RbCs~\cite{BergemanPRA03}. In order to see which 
effect resonant spin-orbit coupling may have on the
stabilization dynamics, we have modified the parameters of the
coupling yielding a potentially favorable and a potentially
detrimental case. While the resulting parametrization may be far from
the true spin-orbit coupling found in the KRb molecule, this approach
allows us to identify the maximum influence that the spin-orbit
coupling may have on the stabilization dynamics. To our surprise, we
found that in both the  potentially favorable and the potentially
detrimental coupling case, the pulse energy required to achieve
population transfer to better than 99\% is significantly increased
compared to the two-state model neglecting the spin-orbit
interaction. This means that the increased size of the state space has
a much larger effect on the optimization than a modification of the
Franck-Condon factors underlying the dynamics. If compared amongst
each other, the potentially detrimental spin-orbit coupling case
requires more pulse energy than the potentially favorable
one. Overall, however, the details of the spin-orbit interaction seem
to have only a minor effect on the stabilization dynamics. 

Third, the spin-orbit interaction in the electronically excited
state allows for population transfer from a singlet-triplet
superposition to a pure singlet level. In order to investigate this as
an optimization problem, our minimal model consists of four electronic
states, the singlet ground and lowest triplet state and the
non-adiabatically coupled electronically excited states. Due to the
symmetry of the electronic states, different polarization components of
the laser field couple to transitions between the singlet and triplet
channels. We found that the further increase in size of the state space as compared
to the three-channel model results in even higher required
pulse energies to achieve population transfer of 99\% or better.
For a shape of the spin-orbit coupling function that is potentially
favorable to the vibrational transfer, we have observed indication
for cooperative behavior between the pulse and the singlet-triplet
transfer due to spin-orbit coupling. We have not seen any evidence for
cooperative behavior in the case of potentially detrimental spin-orbit
coupling. Correspondingly, the required pulse energy is larger for the
potentially detrimental spin-orbit coupling case. 
Since the triplet-singlet transfer is explicitly part of the
optimization problem in the four-state model, it is not surprising
that details of the spin-orbit interaction play a somewhat larger role
than for the three-state model for singlet-to-singlet vibrational
population transfer. 

In summary, independently of the details of the molecular structure,
we have found optimal pulses achieving vibrational population transfer
of KRb molecules to the vibrational ground state with 99\% efficiency
or better. This highlights the power of the optimal control
approach. However, as the complexity of the molecular structure is
increased, the optimal laser fields need to carry more and more pulse
energy. Each individual solution does of course depend on the details
of the molecular structure, and we have analyzed the dynamics under
the optimal pulse in terms of the underlying Franck-Condon maps. 

In the present paper, the initial state was taken to be a highly
excited but pure state. Such a situation is encountered for example if
the molecules are created utilizing a Feshbach
resonance~\cite{OspelkausPRL06}. Transfer to the vibrational
ground state can then be achieved in a purely coherent process where
the pulse absorbs the vibrational excitation energy of the molecule. 
The optimization task becomes more involved if the initial state
corresponds to an incoherent ensemble of vibrationally excited
molecules. Such a situation occurs if the molecules are created by
photoassociation followed by spontaneous emission~\cite{FrancoiseReview}. 
A true cooling scheme is then required where the molecules can dispose
of energy and entropy~\cite{PilletSci08}. While the present study was
confined to vibrational stabilization, it may nevertheless shed light
on the prospects for vibrational cooling. The Franck-Condon map shows
a characteristic parabola whose distribution of weights and tilt
determine whether the probability for heating is larger than that for
cooling or vice versa. Similarly to the case of
LiCs discussed in Ref.~\cite{Faraday}, the Franck-Condon maps
presented above for Na$_2$ and KRb reveal that vibrational cooling by
optical pumping with a spectrally cut femtosecond laser
pulse~\cite{PilletSci08} will not be successful. In order to
preferentially cool instead of heat despite the Franck-Condon map, a
more sophisticated approach than optical pumping would be
required. Here again, optimal control can serve as the tool of
choice. For a toy molecular model, optimally shaped laser pulses
together with spontaneous emission have been predicted to yield a
successful cooling scheme~\cite{AllonCP01}. 

Finally, we point out that we do not obtain an adiabatic passage-like
solution since this is not accessible within our current optimization
approach~\cite{BandJCP94}. In fact, Stimulated Raman Adiabatic passage
and related solutions formally require the infinite time
limit~\cite{HaidongMyPRA10}; and a finite optimization time needs to
be fixed  when numerically solving the optimization
equations. So on one hand optimal control is an extremely convenient 
tool that allows for solving very complex optimization problems. On
the other hand, however, it is not always straightforward to translate
physical considerations such as allowing for the adiabatic limit
into mathematical prescriptions for the 
algorithm. Our future work is therefore dedicated to developing more
versatile optimization algorithms.

\begin{acknowledgments}
  Fabian Borschel has contributed to this work at its initial stage. 
  Financial support  from the Deutsche Forschungsgemeinschaft 
  is gratefully acknowledged.
\end{acknowledgments}


\begin{thebibliography}{50}
\expandafter\ifx\csname natexlab\endcsname\relax\def\natexlab#1{#1}\fi
\expandafter\ifx\csname bibnamefont\endcsname\relax
  \def\bibnamefont#1{#1}\fi
\expandafter\ifx\csname bibfnamefont\endcsname\relax
  \def\bibfnamefont#1{#1}\fi
\expandafter\ifx\csname citenamefont\endcsname\relax
  \def\citenamefont#1{#1}\fi
\expandafter\ifx\csname url\endcsname\relax
  \def\url#1{\texttt{#1}}\fi
\expandafter\ifx\csname urlprefix\endcsname\relax\def\urlprefix{URL }\fi
\providecommand{\bibinfo}[2]{#2}
\providecommand{\eprint}[2][]{\url{#2}}

\bibitem[{\citenamefont{Krems et~al.}(2009)\citenamefont{Krems, Stwalley, and
  Friedrich}}]{ColdMolBook}
\bibinfo{editor}{\bibfnamefont{R.}~\bibnamefont{Krems}},
  \bibinfo{editor}{\bibfnamefont{W.}~\bibnamefont{Stwalley}}, \bibnamefont{and}
  \bibinfo{editor}{\bibfnamefont{B.}~\bibnamefont{Friedrich}}, eds.,
  \emph{\bibinfo{title}{Cold Molecules. Theory, Experiment, Applications}}
  (\bibinfo{publisher}{CRC Press}, \bibinfo{year}{2009}).

\bibitem[{\citenamefont{van~de Meerakker et~al.}(2006)\citenamefont{van~de
  Meerakker, Vanhaecke, and Meijer}}]{MeijerAnnuRevPhysChem06}
\bibinfo{author}{\bibfnamefont{S.~Y.} \bibnamefont{van~de Meerakker}},
  \bibinfo{author}{\bibfnamefont{N.}~\bibnamefont{Vanhaecke}},
  \bibnamefont{and} \bibinfo{author}{\bibfnamefont{G.}~\bibnamefont{Meijer}},
  \bibinfo{journal}{Annu. Rev. Phys. Chem.} \textbf{\bibinfo{volume}{57}},
  \bibinfo{pages}{159} (\bibinfo{year}{2006}).

\bibitem[{\citenamefont{Masnou-Seeuws and Pillet}(2001)}]{FrancoiseReview}
\bibinfo{author}{\bibfnamefont{F.}~\bibnamefont{Masnou-Seeuws}}
  \bibnamefont{and} \bibinfo{author}{\bibfnamefont{P.}~\bibnamefont{Pillet}},
  \bibinfo{journal}{Adv. in At., Mol. and Opt. Phys.}
  \textbf{\bibinfo{volume}{47}}, \bibinfo{pages}{53} (\bibinfo{year}{2001}).

\bibitem[{\citenamefont{Jones et~al.}(2006)\citenamefont{Jones, Tiesinga, Lett,
  and Julienne}}]{JonesRMP06}
\bibinfo{author}{\bibfnamefont{K.~M.} \bibnamefont{Jones}},
  \bibinfo{author}{\bibfnamefont{E.}~\bibnamefont{Tiesinga}},
  \bibinfo{author}{\bibfnamefont{P.~D.} \bibnamefont{Lett}}, \bibnamefont{and}
  \bibinfo{author}{\bibfnamefont{P.~S.} \bibnamefont{Julienne}},
  \bibinfo{journal}{Rev. Mod. Phys.} \textbf{\bibinfo{volume}{78}},
  \bibinfo{pages}{483} (\bibinfo{year}{2006}).

\bibitem[{\citenamefont{K\"{o}hler et~al.}(2006)\citenamefont{K\"{o}hler,
  G\'{o}ral, and Julienne}}]{KoehlerRMP06}
\bibinfo{author}{\bibfnamefont{T.}~\bibnamefont{K\"{o}hler}},
  \bibinfo{author}{\bibfnamefont{K.}~\bibnamefont{G\'{o}ral}},
  \bibnamefont{and} \bibinfo{author}{\bibfnamefont{P.~S.}
  \bibnamefont{Julienne}}, \bibinfo{journal}{Rev. Mod. Phys.}
  \textbf{\bibinfo{volume}{78}}, \bibinfo{eid}{1311}
  (pages~\bibinfo{numpages}{51}) (\bibinfo{year}{2006}).

\bibitem[{\citenamefont{Ni et~al.}(2008)\citenamefont{Ni, Ospelkaus,
  de~Miranda, Pe'er, Neyenhuis, Zirbel, Kotochigova, Julienne, Jin, and
  Ye}}]{NiSci08}
\bibinfo{author}{\bibfnamefont{K.-K.} \bibnamefont{Ni}},
  \bibinfo{author}{\bibfnamefont{S.}~\bibnamefont{Ospelkaus}},
  \bibinfo{author}{\bibfnamefont{M.~H.~G.} \bibnamefont{de~Miranda}},
  \bibinfo{author}{\bibfnamefont{A.}~\bibnamefont{Pe'er}},
  \bibinfo{author}{\bibfnamefont{B.}~\bibnamefont{Neyenhuis}},
  \bibinfo{author}{\bibfnamefont{J.~J.} \bibnamefont{Zirbel}},
  \bibinfo{author}{\bibfnamefont{S.}~\bibnamefont{Kotochigova}},
  \bibinfo{author}{\bibfnamefont{P.~S.} \bibnamefont{Julienne}},
  \bibinfo{author}{\bibfnamefont{D.~S.} \bibnamefont{Jin}}, \bibnamefont{and}
  \bibinfo{author}{\bibfnamefont{J.}~\bibnamefont{Ye}},
  \bibinfo{journal}{Science} \textbf{\bibinfo{volume}{322}},
  \bibinfo{pages}{231} (\bibinfo{year}{2008}).

\bibitem[{\citenamefont{Lang et~al.}(2008)\citenamefont{Lang, Winkler, Strauss,
  Grimm, and Denschlag}}]{LangPRL08}
\bibinfo{author}{\bibfnamefont{F.}~\bibnamefont{Lang}},
  \bibinfo{author}{\bibfnamefont{K.}~\bibnamefont{Winkler}},
  \bibinfo{author}{\bibfnamefont{C.}~\bibnamefont{Strauss}},
  \bibinfo{author}{\bibfnamefont{R.}~\bibnamefont{Grimm}}, \bibnamefont{and}
  \bibinfo{author}{\bibfnamefont{J.~H.} \bibnamefont{Denschlag}},
  \bibinfo{journal}{Phys. Rev. Lett.} \textbf{\bibinfo{volume}{101}},
  \bibinfo{pages}{133005} (\bibinfo{year}{2008}).

\bibitem[{\citenamefont{Sage et~al.}(2005)\citenamefont{Sage, Sainis, Bergeman,
  and DeMille}}]{SagePRL05}
\bibinfo{author}{\bibfnamefont{J.~M.} \bibnamefont{Sage}},
  \bibinfo{author}{\bibfnamefont{S.}~\bibnamefont{Sainis}},
  \bibinfo{author}{\bibfnamefont{T.}~\bibnamefont{Bergeman}}, \bibnamefont{and}
  \bibinfo{author}{\bibfnamefont{D.}~\bibnamefont{DeMille}},
  \bibinfo{journal}{Phys. Rev. Lett.} \textbf{\bibinfo{volume}{94}},
  \bibinfo{pages}{203001} (\bibinfo{year}{2005}).

\bibitem[{\citenamefont{Deiglmayr et~al.}(2008)\citenamefont{Deiglmayr,
  Grochola, Repp, M\"{o}rtlbauer, Gl\"{u}ck, Lange, Dulieu, Wester, and
  Weidem\"{u}ller}}]{DeiglmayrPRL08}
\bibinfo{author}{\bibfnamefont{J.}~\bibnamefont{Deiglmayr}},
  \bibinfo{author}{\bibfnamefont{A.}~\bibnamefont{Grochola}},
  \bibinfo{author}{\bibfnamefont{M.}~\bibnamefont{Repp}},
  \bibinfo{author}{\bibfnamefont{K.}~\bibnamefont{M\"{o}rtlbauer}},
  \bibinfo{author}{\bibfnamefont{C.}~\bibnamefont{Gl\"{u}ck}},
  \bibinfo{author}{\bibfnamefont{J.}~\bibnamefont{Lange}},
  \bibinfo{author}{\bibfnamefont{O.}~\bibnamefont{Dulieu}},
  \bibinfo{author}{\bibfnamefont{R.}~\bibnamefont{Wester}}, \bibnamefont{and}
  \bibinfo{author}{\bibfnamefont{M.}~\bibnamefont{Weidem\"{u}ller}},
  \bibinfo{journal}{Phys. Rev. Lett.} \textbf{\bibinfo{volume}{101}},
  \bibinfo{pages}{133004} (\bibinfo{year}{2008}).

\bibitem[{\citenamefont{Viteau et~al.}(2008)\citenamefont{Viteau, Chotia,
  Allegrini, Bouloufa, Dulieu, Comparat, and Pillet}}]{PilletSci08}
\bibinfo{author}{\bibfnamefont{M.}~\bibnamefont{Viteau}},
  \bibinfo{author}{\bibfnamefont{A.}~\bibnamefont{Chotia}},
  \bibinfo{author}{\bibfnamefont{M.}~\bibnamefont{Allegrini}},
  \bibinfo{author}{\bibfnamefont{N.}~\bibnamefont{Bouloufa}},
  \bibinfo{author}{\bibfnamefont{O.}~\bibnamefont{Dulieu}},
  \bibinfo{author}{\bibfnamefont{D.}~\bibnamefont{Comparat}}, \bibnamefont{and}
  \bibinfo{author}{\bibfnamefont{P.}~\bibnamefont{Pillet}},
  \bibinfo{journal}{Science} \textbf{\bibinfo{volume}{321}},
  \bibinfo{pages}{232} (\bibinfo{year}{2008}).

\bibitem[{\citenamefont{Ospelkaus et~al.}(2010)\citenamefont{Ospelkaus, Ni,
  Qu\'em\'ener, Neyenhuis, Wang, de~Miranda, Bohn, Ye, and
  Jin}}]{OspelkausPRL10}
\bibinfo{author}{\bibfnamefont{S.}~\bibnamefont{Ospelkaus}},
  \bibinfo{author}{\bibfnamefont{K.-K.} \bibnamefont{Ni}},
  \bibinfo{author}{\bibfnamefont{G.}~\bibnamefont{Qu\'em\'ener}},
  \bibinfo{author}{\bibfnamefont{B.}~\bibnamefont{Neyenhuis}},
  \bibinfo{author}{\bibfnamefont{D.}~\bibnamefont{Wang}},
  \bibinfo{author}{\bibfnamefont{M.~H.~G.} \bibnamefont{de~Miranda}},
  \bibinfo{author}{\bibfnamefont{J.~L.} \bibnamefont{Bohn}},
  \bibinfo{author}{\bibfnamefont{J.}~\bibnamefont{Ye}}, \bibnamefont{and}
  \bibinfo{author}{\bibfnamefont{D.~S.} \bibnamefont{Jin}},
  \bibinfo{journal}{Phys. Rev. Lett.} \textbf{\bibinfo{volume}{104}},
  \bibinfo{pages}{030402} (\bibinfo{year}{2010}).

\bibitem[{\citenamefont{Koch et~al.}(2004)\citenamefont{Koch, Palao, Kosloff,
  and Masnou-Seeuws}}]{MyPRA04}
\bibinfo{author}{\bibfnamefont{C.~P.} \bibnamefont{Koch}},
  \bibinfo{author}{\bibfnamefont{J.~P.} \bibnamefont{Palao}},
  \bibinfo{author}{\bibfnamefont{R.}~\bibnamefont{Kosloff}}, \bibnamefont{and}
  \bibinfo{author}{\bibfnamefont{F.}~\bibnamefont{Masnou-Seeuws}},
  \bibinfo{journal}{Phys. Rev. A} \textbf{\bibinfo{volume}{70}},
  \bibinfo{pages}{013402} (\bibinfo{year}{2004}).

\bibitem[{\citenamefont{Roati et~al.}(2002)\citenamefont{Roati, Riboli,
  Modugno, and Inguscio}}]{RoatiPRL02}
\bibinfo{author}{\bibfnamefont{G.}~\bibnamefont{Roati}},
  \bibinfo{author}{\bibfnamefont{F.}~\bibnamefont{Riboli}},
  \bibinfo{author}{\bibfnamefont{G.}~\bibnamefont{Modugno}}, \bibnamefont{and}
  \bibinfo{author}{\bibfnamefont{M.}~\bibnamefont{Inguscio}},
  \bibinfo{journal}{Phys. Rev. Lett.} \textbf{\bibinfo{volume}{89}},
  \bibinfo{pages}{150403} (\bibinfo{year}{2002}).

\bibitem[{\citenamefont{Thalhammer et~al.}(2008)\citenamefont{Thalhammer,
  Barontini, De~Sarlo, Catani, Minardi, and Inguscio}}]{ThalheimerPRL08}
\bibinfo{author}{\bibfnamefont{G.}~\bibnamefont{Thalhammer}},
  \bibinfo{author}{\bibfnamefont{G.}~\bibnamefont{Barontini}},
  \bibinfo{author}{\bibfnamefont{L.}~\bibnamefont{De~Sarlo}},
  \bibinfo{author}{\bibfnamefont{J.}~\bibnamefont{Catani}},
  \bibinfo{author}{\bibfnamefont{F.}~\bibnamefont{Minardi}}, \bibnamefont{and}
  \bibinfo{author}{\bibfnamefont{M.}~\bibnamefont{Inguscio}},
  \bibinfo{journal}{Phys. Rev. Lett.} \textbf{\bibinfo{volume}{100}},
  \bibinfo{pages}{210402} (\bibinfo{year}{2008}).

\bibitem[{\citenamefont{Wang et~al.}(2004)\citenamefont{Wang, Qi, Stone,
  Nikolayeva, Wang, Hattaway, Gensemer, Gould, Eyler, and
  Stwalley}}]{WangPRL04}
\bibinfo{author}{\bibfnamefont{D.}~\bibnamefont{Wang}},
  \bibinfo{author}{\bibfnamefont{J.}~\bibnamefont{Qi}},
  \bibinfo{author}{\bibfnamefont{M.~F.} \bibnamefont{Stone}},
  \bibinfo{author}{\bibfnamefont{O.}~\bibnamefont{Nikolayeva}},
  \bibinfo{author}{\bibfnamefont{H.}~\bibnamefont{Wang}},
  \bibinfo{author}{\bibfnamefont{B.}~\bibnamefont{Hattaway}},
  \bibinfo{author}{\bibfnamefont{S.~D.} \bibnamefont{Gensemer}},
  \bibinfo{author}{\bibfnamefont{P.~L.} \bibnamefont{Gould}},
  \bibinfo{author}{\bibfnamefont{E.~E.} \bibnamefont{Eyler}}, \bibnamefont{and}
  \bibinfo{author}{\bibfnamefont{W.~C.} \bibnamefont{Stwalley}},
  \bibinfo{journal}{Phys. Rev. Lett.} \textbf{\bibinfo{volume}{93}},
  \bibinfo{pages}{243005} (\bibinfo{year}{2004}).

\bibitem[{\citenamefont{Mancini et~al.}(2004)\citenamefont{Mancini, Telles,
  Caires, Bagnato, and Marcassa}}]{ManciniPRL04}
\bibinfo{author}{\bibfnamefont{M.~W.} \bibnamefont{Mancini}},
  \bibinfo{author}{\bibfnamefont{G.~D.} \bibnamefont{Telles}},
  \bibinfo{author}{\bibfnamefont{A.~R.~L.} \bibnamefont{Caires}},
  \bibinfo{author}{\bibfnamefont{V.~S.} \bibnamefont{Bagnato}},
  \bibnamefont{and} \bibinfo{author}{\bibfnamefont{L.~G.}
  \bibnamefont{Marcassa}}, \bibinfo{journal}{Phys. Rev. Lett.}
  \textbf{\bibinfo{volume}{92}}, \bibinfo{pages}{133203}
  (\bibinfo{year}{2004}).

\bibitem[{\citenamefont{Ospelkaus et~al.}(2006)\citenamefont{Ospelkaus,
  Ospelkaus, Humbert, Ernst, Sengstock, and Bongs}}]{OspelkausPRL06}
\bibinfo{author}{\bibfnamefont{C.}~\bibnamefont{Ospelkaus}},
  \bibinfo{author}{\bibfnamefont{S.}~\bibnamefont{Ospelkaus}},
  \bibinfo{author}{\bibfnamefont{L.}~\bibnamefont{Humbert}},
  \bibinfo{author}{\bibfnamefont{P.}~\bibnamefont{Ernst}},
  \bibinfo{author}{\bibfnamefont{K.}~\bibnamefont{Sengstock}},
  \bibnamefont{and} \bibinfo{author}{\bibfnamefont{K.}~\bibnamefont{Bongs}},
  \bibinfo{journal}{Phys. Rev. Lett.} \textbf{\bibinfo{volume}{97}},
  \bibinfo{pages}{120402} (\bibinfo{year}{2006}).

\bibitem[{\citenamefont{Klempt et~al.}(2008)\citenamefont{Klempt, Henninger,
  Topic, Scherer, Kattner, Tiemann, Ertmer, and Arlt}}]{KlemptPRA08}
\bibinfo{author}{\bibfnamefont{C.}~\bibnamefont{Klempt}},
  \bibinfo{author}{\bibfnamefont{T.}~\bibnamefont{Henninger}},
  \bibinfo{author}{\bibfnamefont{O.}~\bibnamefont{Topic}},
  \bibinfo{author}{\bibfnamefont{M.}~\bibnamefont{Scherer}},
  \bibinfo{author}{\bibfnamefont{L.}~\bibnamefont{Kattner}},
  \bibinfo{author}{\bibfnamefont{E.}~\bibnamefont{Tiemann}},
  \bibinfo{author}{\bibfnamefont{W.}~\bibnamefont{Ertmer}}, \bibnamefont{and}
  \bibinfo{author}{\bibfnamefont{J.~J.} \bibnamefont{Arlt}},
  \bibinfo{journal}{Phys. Rev. A} \textbf{\bibinfo{volume}{78}},
  \bibinfo{pages}{061602} (\bibinfo{year}{2008}).

\bibitem[{\citenamefont{Aikawa et~al.}(2009)\citenamefont{Aikawa, Akamatsu,
  Kobayashi, Ueda, Kishimoto, and Inouye}}]{AikawaNJP09}
\bibinfo{author}{\bibfnamefont{K.}~\bibnamefont{Aikawa}},
  \bibinfo{author}{\bibfnamefont{D.}~\bibnamefont{Akamatsu}},
  \bibinfo{author}{\bibfnamefont{J.}~\bibnamefont{Kobayashi}},
  \bibinfo{author}{\bibfnamefont{M.}~\bibnamefont{Ueda}},
  \bibinfo{author}{\bibfnamefont{T.}~\bibnamefont{Kishimoto}},
  \bibnamefont{and} \bibinfo{author}{\bibfnamefont{S.}~\bibnamefont{Inouye}},
  \bibinfo{journal}{New J. Phys.} \textbf{\bibinfo{volume}{11}},
  \bibinfo{pages}{055035} (\bibinfo{year}{2009}).

\bibitem[{\citenamefont{Kim et~al.}(2009)\citenamefont{Kim, Wang, Eyler, Gould,
  and Stwalley}}]{KimNJP09}
\bibinfo{author}{\bibfnamefont{J.~T.} \bibnamefont{Kim}},
  \bibinfo{author}{\bibfnamefont{D.}~\bibnamefont{Wang}},
  \bibinfo{author}{\bibfnamefont{E.~E.} \bibnamefont{Eyler}},
  \bibinfo{author}{\bibfnamefont{P.~L.} \bibnamefont{Gould}}, \bibnamefont{and}
  \bibinfo{author}{\bibfnamefont{W.~C.} \bibnamefont{Stwalley}},
  \bibinfo{journal}{New J. Phys.} \textbf{\bibinfo{volume}{11}},
  \bibinfo{pages}{055020} (\bibinfo{year}{2009}).

\bibitem[{\citenamefont{Amiot et~al.}(1999)\citenamefont{Amiot, Dulieu, and
  Verg\`{e}s}}]{AmiotPRL99}
\bibinfo{author}{\bibfnamefont{C.}~\bibnamefont{Amiot}},
  \bibinfo{author}{\bibfnamefont{O.}~\bibnamefont{Dulieu}}, \bibnamefont{and}
  \bibinfo{author}{\bibfnamefont{J.}~\bibnamefont{Verg\`{e}s}},
  \bibinfo{journal}{Phys. Rev. Lett.} \textbf{\bibinfo{volume}{83}},
  \bibinfo{pages}{2316} (\bibinfo{year}{1999}).

\bibitem[{\citenamefont{Pechkis et~al.}(2007)\citenamefont{Pechkis, Wang,
  Huang, Eyler, Gould, Stwalley, and Koch}}]{HyewonMyPRA07}
\bibinfo{author}{\bibfnamefont{H.~K.} \bibnamefont{Pechkis}},
  \bibinfo{author}{\bibfnamefont{D.}~\bibnamefont{Wang}},
  \bibinfo{author}{\bibfnamefont{Y.}~\bibnamefont{Huang}},
  \bibinfo{author}{\bibfnamefont{E.~E.} \bibnamefont{Eyler}},
  \bibinfo{author}{\bibfnamefont{P.~L.} \bibnamefont{Gould}},
  \bibinfo{author}{\bibfnamefont{W.~C.} \bibnamefont{Stwalley}},
  \bibnamefont{and} \bibinfo{author}{\bibfnamefont{C.~P.} \bibnamefont{Koch}},
  \bibinfo{journal}{Phys. Rev. A} \textbf{\bibinfo{volume}{76}},
  \bibinfo{pages}{022504} (\bibinfo{year}{2007}).

\bibitem[{\citenamefont{Ghosal et~al.}(2009)\citenamefont{Ghosal, Doyle, Koch,
  and Hutson}}]{GhosalMyNJP09}
\bibinfo{author}{\bibfnamefont{S.}~\bibnamefont{Ghosal}},
  \bibinfo{author}{\bibfnamefont{R.~J.} \bibnamefont{Doyle}},
  \bibinfo{author}{\bibfnamefont{C.~P.} \bibnamefont{Koch}}, \bibnamefont{and}
  \bibinfo{author}{\bibfnamefont{J.~M.} \bibnamefont{Hutson}},
  \bibinfo{journal}{New J. Phys.} \textbf{\bibinfo{volume}{11}},
  \bibinfo{pages}{055011} (\bibinfo{year}{2009}).

\bibitem[{\citenamefont{Koch and Kosloff}(2009)}]{MyPRL09}
\bibinfo{author}{\bibfnamefont{C.~P.} \bibnamefont{Koch}} \bibnamefont{and}
  \bibinfo{author}{\bibfnamefont{R.}~\bibnamefont{Kosloff}},
  \bibinfo{journal}{Phys. Rev. Lett.} \textbf{\bibinfo{volume}{103}},
  \bibinfo{pages}{260401} (\bibinfo{year}{2009}),
  \bibinfo{note}{arXiv:0905.3251}.

\bibitem[{\citenamefont{Naidon and Masnou-Seeuws}(2003)}]{PascalPRA03}
\bibinfo{author}{\bibfnamefont{P.}~\bibnamefont{Naidon}} \bibnamefont{and}
  \bibinfo{author}{\bibfnamefont{F.}~\bibnamefont{Masnou-Seeuws}},
  \bibinfo{journal}{Phys. Rev. A} \textbf{\bibinfo{volume}{68}},
  \bibinfo{pages}{033612} (\bibinfo{year}{2003}).

\bibitem[{\citenamefont{Kosloff}(1994)}]{RonnieReview94}
\bibinfo{author}{\bibfnamefont{R.}~\bibnamefont{Kosloff}},
  \bibinfo{journal}{Annu. Rev. Phys. Chem.} \textbf{\bibinfo{volume}{45}},
  \bibinfo{pages}{145} (\bibinfo{year}{1994}).

\bibitem[{\citenamefont{Kokoouline et~al.}(1999)\citenamefont{Kokoouline,
  Dulieu, Kosloff, and Masnou-Seeuws}}]{SlavaJCP99}
\bibinfo{author}{\bibfnamefont{V.}~\bibnamefont{Kokoouline}},
  \bibinfo{author}{\bibfnamefont{O.}~\bibnamefont{Dulieu}},
  \bibinfo{author}{\bibfnamefont{R.}~\bibnamefont{Kosloff}}, \bibnamefont{and}
  \bibinfo{author}{\bibfnamefont{F.}~\bibnamefont{Masnou-Seeuws}},
  \bibinfo{journal}{J. Chem. Phys.} \textbf{\bibinfo{volume}{110}},
  \bibinfo{pages}{9865} (\bibinfo{year}{1999}).

\bibitem[{\citenamefont{Willner et~al.}(2004)\citenamefont{Willner, Dulieu, and
  Masnou-Seeuws}}]{WillnerJCP04}
\bibinfo{author}{\bibfnamefont{K.}~\bibnamefont{Willner}},
  \bibinfo{author}{\bibfnamefont{O.}~\bibnamefont{Dulieu}}, \bibnamefont{and}
  \bibinfo{author}{\bibfnamefont{F.}~\bibnamefont{Masnou-Seeuws}},
  \bibinfo{journal}{J. Chem. Phys.} \textbf{\bibinfo{volume}{120}},
  \bibinfo{pages}{548} (\bibinfo{year}{2004}).

\bibitem[{\citenamefont{Kallush and Kosloff}(2006)}]{ShimshonCPL06}
\bibinfo{author}{\bibfnamefont{S.}~\bibnamefont{Kallush}} \bibnamefont{and}
  \bibinfo{author}{\bibfnamefont{R.}~\bibnamefont{Kosloff}},
  \bibinfo{journal}{Chem. Phys. Lett.} \textbf{\bibinfo{volume}{433}},
  \bibinfo{pages}{221} (\bibinfo{year}{2006}).

\bibitem[{\citenamefont{Palao and Kosloff}(2003)}]{JosePRA03}
\bibinfo{author}{\bibfnamefont{J.~P.} \bibnamefont{Palao}} \bibnamefont{and}
  \bibinfo{author}{\bibfnamefont{R.}~\bibnamefont{Kosloff}},
  \bibinfo{journal}{Phys. Rev. A} \textbf{\bibinfo{volume}{68}},
  \bibinfo{pages}{062308} (\bibinfo{year}{2003}).

\bibitem[{\citenamefont{Krotov}(2008)}]{KrotovDok08}
\bibinfo{author}{\bibfnamefont{V.~F.} \bibnamefont{Krotov}},
  \bibinfo{journal}{Doklady Mathematics} \textbf{\bibinfo{volume}{78}},
  \bibinfo{pages}{949} (\bibinfo{year}{2008}).

\bibitem[{\citenamefont{Krotov}(2009)}]{KrotovAC09}
\bibinfo{author}{\bibfnamefont{V.~F.} \bibnamefont{Krotov}},
  \bibinfo{journal}{Automation and Remote Control}
  \textbf{\bibinfo{volume}{70}}, \bibinfo{pages}{357} (\bibinfo{year}{2009}).

\bibitem[{\citenamefont{Sklarz and Tannor}(2002)}]{SklarzPRA02}
\bibinfo{author}{\bibfnamefont{S.~E.} \bibnamefont{Sklarz}} \bibnamefont{and}
  \bibinfo{author}{\bibfnamefont{D.~J.} \bibnamefont{Tannor}},
  \bibinfo{journal}{Phys. Rev. A} \textbf{\bibinfo{volume}{66}},
  \bibinfo{pages}{053619} (\bibinfo{year}{2002}).

\bibitem[{\citenamefont{Konnov and Krotov}(1999)}]{KonnovAC99}
\bibinfo{author}{\bibfnamefont{A.~I.} \bibnamefont{Konnov}} \bibnamefont{and}
  \bibinfo{author}{\bibfnamefont{V.~F.} \bibnamefont{Krotov}},
  \bibinfo{journal}{Automation and Remote Control}
  \textbf{\bibinfo{volume}{60}}, \bibinfo{pages}{1427} (\bibinfo{year}{1999}).

\bibitem[{\citenamefont{Samuelis et~al.}(2000)\citenamefont{Samuelis, Tiesinga,
  Laue, Elbs, Kn\"ockel, and Tiemann}}]{SamuelisPRA00}
\bibinfo{author}{\bibfnamefont{C.}~\bibnamefont{Samuelis}},
  \bibinfo{author}{\bibfnamefont{E.}~\bibnamefont{Tiesinga}},
  \bibinfo{author}{\bibfnamefont{T.}~\bibnamefont{Laue}},
  \bibinfo{author}{\bibfnamefont{M.}~\bibnamefont{Elbs}},
  \bibinfo{author}{\bibfnamefont{H.}~\bibnamefont{Kn\"ockel}},
  \bibnamefont{and} \bibinfo{author}{\bibfnamefont{E.}~\bibnamefont{Tiemann}},
  \bibinfo{journal}{Phys. Rev. A} \textbf{\bibinfo{volume}{63}},
  \bibinfo{pages}{012710} (\bibinfo{year}{2000}).

\bibitem[{\citenamefont{Tiemann et~al.}(1996)\citenamefont{Tiemann, K\"ockel,
  and Richling}}]{TiemannZfPD96}
\bibinfo{author}{\bibfnamefont{E.}~\bibnamefont{Tiemann}},
  \bibinfo{author}{\bibfnamefont{H.}~\bibnamefont{K\"ockel}}, \bibnamefont{and}
  \bibinfo{author}{\bibfnamefont{H.}~\bibnamefont{Richling}},
  \bibinfo{journal}{Z. Phys. D.} \textbf{\bibinfo{volume}{37}},
  \bibinfo{pages}{323} (\bibinfo{year}{1996}).

\bibitem[{\citenamefont{Pashov et~al.}(2007)\citenamefont{Pashov, Docenko,
  Tamanis, Ferber, Kn\"{o}ckel, and Tiemann}}]{PashovPRA07}
\bibinfo{author}{\bibfnamefont{A.}~\bibnamefont{Pashov}},
  \bibinfo{author}{\bibfnamefont{O.}~\bibnamefont{Docenko}},
  \bibinfo{author}{\bibfnamefont{M.}~\bibnamefont{Tamanis}},
  \bibinfo{author}{\bibfnamefont{R.}~\bibnamefont{Ferber}},
  \bibinfo{author}{\bibfnamefont{H.}~\bibnamefont{Kn\"{o}ckel}},
  \bibnamefont{and} \bibinfo{author}{\bibfnamefont{E.}~\bibnamefont{Tiemann}},
  \bibinfo{journal}{Phys. Rev. A} \textbf{\bibinfo{volume}{76}},
  \bibinfo{pages}{022511} (\bibinfo{year}{2007}).

\bibitem[{\citenamefont{Rousseau et~al.}(2000)\citenamefont{Rousseau, Allouche,
  and Aubert-Frécon}}]{RousseauJMolSpec00}
\bibinfo{author}{\bibfnamefont{S.}~\bibnamefont{Rousseau}},
  \bibinfo{author}{\bibfnamefont{A.}~\bibnamefont{Allouche}}, \bibnamefont{and}
  \bibinfo{author}{\bibfnamefont{M.}~\bibnamefont{Aubert-Frécon}},
  \bibinfo{journal}{Journal of Mol. Spectrosc.} \textbf{\bibinfo{volume}{203}},
  \bibinfo{pages}{235} (\bibinfo{year}{2000}), ISSN \bibinfo{issn}{0022-2852}.

\bibitem[{\citenamefont{Marinescu and Sadeghpour}(1999)}]{MarinescuPRA99}
\bibinfo{author}{\bibfnamefont{M.}~\bibnamefont{Marinescu}} \bibnamefont{and}
  \bibinfo{author}{\bibfnamefont{H.~R.} \bibnamefont{Sadeghpour}},
  \bibinfo{journal}{Phys. Rev. A} \textbf{\bibinfo{volume}{59}},
  \bibinfo{pages}{390} (\bibinfo{year}{1999}).

\bibitem[{\citenamefont{Kallush and Kosloff}(2007)}]{KallushPRA07}
\bibinfo{author}{\bibfnamefont{S.}~\bibnamefont{Kallush}} \bibnamefont{and}
  \bibinfo{author}{\bibfnamefont{R.}~\bibnamefont{Kosloff}},
  \bibinfo{journal}{Phys. Rev. A} \textbf{\bibinfo{volume}{76}},
  \bibinfo{pages}{053408} (\bibinfo{year}{2007}).

\bibitem[{\citenamefont{Lapert et~al.}(2009)\citenamefont{Lapert, Tehini,
  Turinici, and Sugny}}]{LapertPRA09}
\bibinfo{author}{\bibfnamefont{M.}~\bibnamefont{Lapert}},
  \bibinfo{author}{\bibfnamefont{R.}~\bibnamefont{Tehini}},
  \bibinfo{author}{\bibfnamefont{G.}~\bibnamefont{Turinici}}, \bibnamefont{and}
  \bibinfo{author}{\bibfnamefont{D.}~\bibnamefont{Sugny}},
  \bibinfo{journal}{Phys. Rev. A} \textbf{\bibinfo{volume}{79}},
  \bibinfo{pages}{063411} (\bibinfo{year}{2009}).

\bibitem[{\citenamefont{Schr\"oder and Brown}(2009)}]{SchroederNJP09}
\bibinfo{author}{\bibfnamefont{M.}~\bibnamefont{Schr\"oder}} \bibnamefont{and}
  \bibinfo{author}{\bibfnamefont{A.}~\bibnamefont{Brown}},
  \bibinfo{journal}{New J. Phys.} \textbf{\bibinfo{volume}{11}},
  \bibinfo{pages}{105031 (13pp)} (\bibinfo{year}{2009}).

\bibitem[{\citenamefont{Bergeman et~al.}(2006)\citenamefont{Bergeman, Qi, Wang,
  Huang, Pechkis, Eyler, Gould, Stwalley, Cline, Miller
  et~al.}}]{BergemanJPB06}
\bibinfo{author}{\bibfnamefont{T.}~\bibnamefont{Bergeman}},
  \bibinfo{author}{\bibfnamefont{J.}~\bibnamefont{Qi}},
  \bibinfo{author}{\bibfnamefont{D.}~\bibnamefont{Wang}},
  \bibinfo{author}{\bibfnamefont{Y.}~\bibnamefont{Huang}},
  \bibinfo{author}{\bibfnamefont{H.~K.} \bibnamefont{Pechkis}},
  \bibinfo{author}{\bibfnamefont{E.~E.} \bibnamefont{Eyler}},
  \bibinfo{author}{\bibfnamefont{P.~L.} \bibnamefont{Gould}},
  \bibinfo{author}{\bibfnamefont{W.~C.} \bibnamefont{Stwalley}},
  \bibinfo{author}{\bibfnamefont{R.~A.} \bibnamefont{Cline}},
  \bibinfo{author}{\bibfnamefont{J.~D.} \bibnamefont{Miller}},
  \bibnamefont{et~al.}, \bibinfo{journal}{J. Phys. B}
  \textbf{\bibinfo{volume}{39}}, \bibinfo{pages}{S813} (\bibinfo{year}{2006}).

\bibitem[{\citenamefont{Bergeman et~al.}(2003)\citenamefont{Bergeman, Fellows,
  Gutterres, and Amiot}}]{BergemanPRA03}
\bibinfo{author}{\bibfnamefont{T.}~\bibnamefont{Bergeman}},
  \bibinfo{author}{\bibfnamefont{C.~E.} \bibnamefont{Fellows}},
  \bibinfo{author}{\bibfnamefont{R.~F.} \bibnamefont{Gutterres}},
  \bibnamefont{and} \bibinfo{author}{\bibfnamefont{C.}~\bibnamefont{Amiot}},
  \bibinfo{journal}{Phys. Rev. A} \textbf{\bibinfo{volume}{67}},
  \bibinfo{pages}{050501} (\bibinfo{year}{2003}).

\bibitem[{\citenamefont{Koch et~al.}(2006)\citenamefont{Koch, Kosloff, and
  Masnou-Seeuws}}]{MyPRA06b}
\bibinfo{author}{\bibfnamefont{C.~P.} \bibnamefont{Koch}},
  \bibinfo{author}{\bibfnamefont{R.}~\bibnamefont{Kosloff}}, \bibnamefont{and}
  \bibinfo{author}{\bibfnamefont{F.}~\bibnamefont{Masnou-Seeuws}},
  \bibinfo{journal}{Phys. Rev. A} \textbf{\bibinfo{volume}{73}},
  \bibinfo{pages}{043409} (\bibinfo{year}{2006}).

\bibitem[{\citenamefont{Caneva et~al.}(2009)\citenamefont{Caneva, Murphy,
  Calarco, Fazio, Montangero, Giovannetti, and Santoro}}]{CanevaPRL10}
\bibinfo{author}{\bibfnamefont{T.}~\bibnamefont{Caneva}},
  \bibinfo{author}{\bibfnamefont{M.}~\bibnamefont{Murphy}},
  \bibinfo{author}{\bibfnamefont{T.}~\bibnamefont{Calarco}},
  \bibinfo{author}{\bibfnamefont{R.}~\bibnamefont{Fazio}},
  \bibinfo{author}{\bibfnamefont{S.}~\bibnamefont{Montangero}},
  \bibinfo{author}{\bibfnamefont{V.}~\bibnamefont{Giovannetti}},
  \bibnamefont{and} \bibinfo{author}{\bibfnamefont{G.~E.}
  \bibnamefont{Santoro}}, \bibinfo{journal}{Phys. Rev. Lett.}
  \textbf{\bibinfo{volume}{103}}, \bibinfo{pages}{240501}
  (\bibinfo{year}{2009}).

\bibitem[{\citenamefont{Hornung et~al.}(2009)\citenamefont{Hornung,
      Motzkus, and de Vivie-Riedle}}]{HornungJCP01}
\bibinfo{author}{\bibfnamefont{T.}~\bibnamefont{Hornung}},
  \bibinfo{author}{\bibfnamefont{M.}~\bibnamefont{Motzkus}},
  \bibnamefont{and} \bibinfo{author}{\bibfnamefont{R.}
  \bibnamefont{de Vivie-Riedle}}, \bibinfo{journal}{J. Chem. Phys.}
  \textbf{\bibinfo{volume}{115}}, \bibinfo{pages}{3105}
  (\bibinfo{year}{2001}).

\bibitem[{\citenamefont{Bayer et~al.}(2009)\citenamefont{Baier, Wollenhaupt,
      Sarpe-Tudoran, and Baumert}}]{BayerPRL09}
\bibinfo{author}{\bibfnamefont{T.}~\bibnamefont{Baier}},
  \bibinfo{author}{\bibfnamefont{M.}~\bibnamefont{Wollenhaupt}},
  \bibinfo{author}{\bibfnamefont{C.}~\bibnamefont{Sarpe-Tudoran}},
  \bibnamefont{and} \bibinfo{author}{\bibfnamefont{T.}
  \bibnamefont{Baumert}}, \bibinfo{journal}{Phys. Rev. Lett.}
  \textbf{\bibinfo{volume}{102}}, \bibinfo{pages}{023004}
  (\bibinfo{year}{2009}).

\bibitem[{\citenamefont{Skinner et~al.}(2003)\citenamefont{Skinner, Reiss,
      Luy, Khaneja, and Glaser}}]{SkinnerJMR03}
\bibinfo{author}{\bibfnamefont{T.~E.}~\bibnamefont{Skinner}},
  \bibinfo{author}{\bibfnamefont{T.}~\bibnamefont{Reiss}},
  \bibinfo{author}{\bibfnamefont{B.}~\bibnamefont{Luy}},
  \bibinfo{author}{\bibfnamefont{N.}~\bibnamefont{Khaneja}},
  \bibnamefont{and} \bibinfo{author}{\bibfnamefont{S.~J.}
  \bibnamefont{Glaser}}, \bibinfo{journal}{J. Magn. Reson.}
  \textbf{\bibinfo{volume}{163}}, \bibinfo{pages}{8}
  (\bibinfo{year}{2003}).
  
\bibitem[{\citenamefont{Kosloff et~al.}(2009)\citenamefont{Kosloff, Pillet,
  Koch, Ye, Barker, Deiglmayr, and Pichler}}]{Faraday}
\bibinfo{author}{\bibfnamefont{R.}~\bibnamefont{Kosloff}},
  \bibinfo{author}{\bibfnamefont{P.}~\bibnamefont{Pillet}},
  \bibinfo{author}{\bibfnamefont{C.~P.}~\bibnamefont{Koch}},
  \bibinfo{author}{\bibfnamefont{J.}~\bibnamefont{Ye}},
  \bibinfo{author}{\bibfnamefont{P.}~\bibnamefont{Barker}},
  \bibinfo{author}{\bibfnamefont{J.}~\bibnamefont{Deiglmayr}},
  \bibnamefont{and} \bibinfo{author}{\bibfnamefont{M.}~\bibnamefont{Pichler}},
  \bibinfo{journal}{Faraday Discuss.} \textbf{\bibinfo{volume}{142}},
  \bibinfo{pages}{319} (\bibinfo{year}{2009}).

\bibitem[{\citenamefont{Bartana et~al.}(2001)\citenamefont{Bartana, Kosloff,
  and Tannor}}]{AllonCP01}
\bibinfo{author}{\bibfnamefont{A.}~\bibnamefont{Bartana}},
  \bibinfo{author}{\bibfnamefont{R.}~\bibnamefont{Kosloff}}, \bibnamefont{and}
  \bibinfo{author}{\bibfnamefont{D.~J.} \bibnamefont{Tannor}},
  \bibinfo{journal}{Chem. Phys.} \textbf{\bibinfo{volume}{267}},
  \bibinfo{pages}{195} (\bibinfo{year}{2001}).

\bibitem[{\citenamefont{Band and Magnes}(1994)}]{BandJCP94}
\bibinfo{author}{\bibfnamefont{Y.~B.} \bibnamefont{Band}} \bibnamefont{and}
  \bibinfo{author}{\bibfnamefont{O.}~\bibnamefont{Magnes}},
  \bibinfo{journal}{J. Chem. Phys.} \textbf{\bibinfo{volume}{101}},
  \bibinfo{pages}{7528} (\bibinfo{year}{1994}).

\bibitem[{\citenamefont{Yuan et~al.}(2010)\citenamefont{Yuan, Koch, Salamon,
  and Tannor}}]{HaidongMyPRA10}
\bibinfo{author}{\bibfnamefont{H.}~\bibnamefont{Yuan}},
  \bibinfo{author}{\bibfnamefont{C.~P.} \bibnamefont{Koch}},
  \bibinfo{author}{\bibfnamefont{P.}~\bibnamefont{Salamon}}, \bibnamefont{and}
  \bibinfo{author}{\bibfnamefont{D.~J.} \bibnamefont{Tannor}},
  \bibinfo{journal}{arXiv:1004.4050v1}  (\bibinfo{year}{2010}).

\end{thebibliography}

\end{document}